\documentclass[%
superscriptaddress,
twocolumn,
amsmath,amssymb,
aps,
prb,
floatfix
]{revtex4-2}

\usepackage{dcolumn}
\usepackage{bm}
\usepackage[paperwidth=210mm,paperheight=297mm,centering,hmargin=2.3cm,tmargin=2.8cm,bmargin=3.4cm]{geometry}
\usepackage{booktabs}
\usepackage{subfigure}
\usepackage{graphicx,epsf}
\usepackage{amsfonts}
\usepackage{amssymb}
\usepackage{amsmath}
\usepackage{overpic}
\usepackage{braket}
\usepackage{xcolor}
\usepackage{cancel}
\usepackage{fancyhdr}
\usepackage[normalem]{ulem}
\makeatletter
\def\maketitle{
\@author@finish
\title@column\titleblock@produce
\suppressfloats[t]}
\makeatother

\begin{document}

\title{Magnetic edge fields in UTe$_2$ near zero background fields
}

\author{Yusuke Iguchi}
\affiliation{Geballe Laboratory for Advanced Materials, Stanford University, Stanford, California 94305, USA}
\affiliation{Stanford Institute for Materials and Energy Sciences, SLAC National Accelerator Laboratory, 2575 Sand Hill Road, Menlo Park, California 94025, USA}

\author{Huiyuan Man}
\affiliation{Geballe Laboratory for Advanced Materials, Stanford University, Stanford, California 94305, USA}
\affiliation{Stanford Nano Shared Facilities, Stanford University, Stanford, CA 94305, USA}

\author{S. M. Thomas}
\affiliation{Los Alamos National Laboratory, Los Alamos, New Mexico 87545, USA}

\author{Filip Ronning}
\affiliation{Los Alamos National Laboratory, Los Alamos, New Mexico 87545, USA}

\author{Jun Ishizuka}
\affiliation{Faculty of Engineering, Niigata University, Ikarashi, Niigata 950-2181, Japan}
\affiliation{Institute for Theoretical Physics, ETH Z\"{u}rich, 8093 Z\"{u}rich, Switzerland}

\author{Manfred Sigrist}
\affiliation{Institute for Theoretical Physics, ETH Z\"{u}rich, 8093 Z\"{u}rich, Switzerland}

\author{Priscila F. S. Rosa}
\affiliation{Los Alamos National Laboratory, Los Alamos, New Mexico 87545, USA}

\author{Kathryn A. Moler} 
\affiliation{Geballe Laboratory for Advanced Materials, Stanford University, Stanford, California 94305, USA}
\affiliation{Stanford Institute for Materials and Energy Sciences, SLAC National Accelerator Laboratory, 2575 Sand Hill Road, Menlo Park, California 94025, USA}
\affiliation{Department of Applied Physics, Stanford University, Stanford, California 94305, USA}

\begin{abstract}
Chiral superconductors are theorized to exhibit spontaneous edge currents. Here, we found magnetic fields at the edges of UTe$_2$, a candidate odd-parity chiral superconductor, that seem to agree with predictions for a chiral order parameter. However, we did not detect the chiral domains that would be expected, and recent polar Kerr and muon spin relaxation data in nominally clean samples argue against chiral superconductivity. Our results show that hidden sources of magnetism must be carefully ruled out when using spontaneous edge currents to identify chiral superconductivity.  
\end{abstract}

\maketitle



\section{Introduction}
Chiral superconductors are intriguing quantum materials whose complex superconducting gap functions break time-reversal symmetry (TRS) and mirror symmetry. A chiral superconducting state may become energetically favorable by eliminating nodes in the gap structure, and a prototypical example is the odd-parity chiral {\it p}-wave gap function ($k_x \pm i k_y$). Zero-energy excitation edge modes, such as spontaneous edge currents or Majorana fermions~\cite{Tanaka2012,Alicea2012NewSystems}, are predicted to emerge in topologically nontrivial chiral superconductors \cite{Mizushima2015Symmetry3He-B,Nie2020EdgeRevisited}. Theoretical models also predict that chiral edge currents induce a spontaneous magnetic field at the sample edge or the boundary of chiral domains~\cite{Matsumoto1999QuasiparticleSuperconductor}. Recently, chiral domain structures were experimentally visualized in chiral {\it p}-wave superfluid $^3$He-A using a specialized magnetic resonance imaging technique~\cite{Kasai2018Chiral}.

Various methods to probe TRS breaking in superconductors have been developed, such as muon spin relaxation ($\mu$SR), polar Kerr rotation, scanning Hall probe microscopy, and superconducting quantum interference device (SQUID) microscopy~\cite{Kallin2016,Wysokinski2019}. More recently, small-angle neutron scattering has also been employed to detect TRS breaking~\cite{Avers2020Broken}. Materials hosting chiral superconductivity, however, remain rare. For example, $\mu$SR or polar Kerr rotation measurements detected TRS breaking in UPt$_3$~\cite{Luke1993,Schemm2014}, Sr$_2$RuO$_4$~\cite{Luke1998,Xia2006}, PrOs$_4$Sb$_{12}$~\cite{Aoki2003,Levenson2018}, and URu$_2$Si$_2$~\cite{Schemm2015}. Nonetheless, no direct visualization of chiral edge currents has been reported in any chiral superconductor candidate to date~\cite{Bjornsson2005,Kirtley2007Upper,Hicks2010Limits,Curran2014,Iguchi2021,Curran2023}. 

Scanning SQUID microscopy is the ideal tool to probe chiral domains because noise levels are expected to be smaller than spontaneous fields estimated by theoretical calculations (see Fig. 1 and, for example, calculations by Matsumoto and Sigrist for a $k_x \pm i k_y$ state  ~\cite{Matsumoto1999QuasiparticleSuperconductor}). In fact, the upper limit of spontaneous magnetization or the size of possible chiral domain structures has been estimated for Sr$_2$RuO$_4$~\cite{Kirtley2007Upper,Hicks2010Limits,Curran2014,Curran2023}, PrOs$_4$Sb$_{12}$~\cite{Hicks2010Limits}, and URu$_2$Si$_2$~\cite{Iguchi2021}.

\begin{figure*}[tb]
\begin{center}
\includegraphics*[width=14cm]{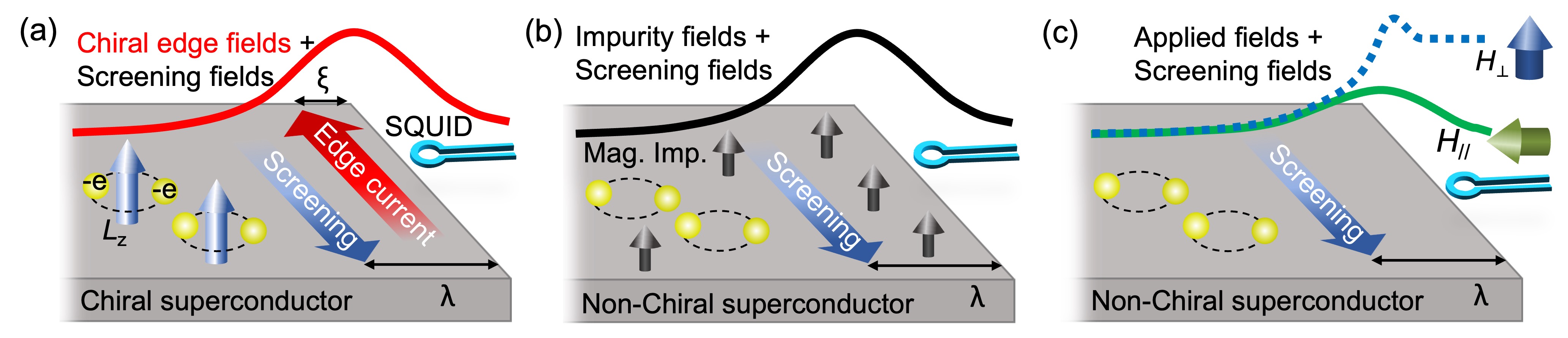}
\caption{{ Scanning SQUID Magnetometry on chiral superconductors.} (a) Schematic of spontaneous magnetic fields in a chiral superconductor at zero background fields~\cite{Hicks2010Limits}, where $L_z$ is angular momentum and $e$ is the elementary charge. This field results from the partial cancellation of chiral edge currents by Meissner screening currents, characterized by the coherence length $\xi$ and the magnetic penetration depth $\lambda$, respectively. (b) Meissner screening fields in a non-chiral superconductor with magnetic impurities, the magnetic moments of which are polarized along the out-of-plane direction~\cite{Bluhm2007Magnetic}. (c) Meissner screening fields in a non-chiral superconductor at applied background fields.}
\end{center}
\end{figure*}

Here, we employ scanning SQUID microscopy to investigate 
recently-discovered odd-parity superconductor UTe$_2$, which has been put forward as a candidate for chiral superconductivity\cite{Ran2019Nearly,Aoki2022Unconventional}.
Evidence for odd-parity superconductivity stems from nuclear magnetic resonance (NMR) Knight shift measurements and upper critical fields above the weak-coupling Pauli limit \cite{Nakamine2019Sup,Nakamine2021Ani,Matsumura2023Large}. The existence of point nodes in the superconducting gap structure is supported by thermodynamic \cite{Ran2019Nearly,Metz2019Point,Kittaka2020Ori,Lee2023Ani} and superfluid density measurements \cite{Bae2021Ano,Ishihara2021Chiral,Iguchi2023Microscopic}. 
 
Evidence for chiral superconductivity was also suggested by various experiments~\cite{Jiao2020Chiral,Hayes2021Multi,Ishihara2021Chiral,Wei2022}. For example, scanning tunneling spectroscopy on the step edge of a (01$\bar{1}$) plane detected an asymmetry in the differential tunneling conductance spectrum depending on the edge direction \cite{Jiao2020Chiral}. Polar Kerr rotation measurements observed a finite Kerr angle that was trainable with $c$-axis magnetic field, which suggests a chiral superconducting order parameter $B_{3u}+iB_{2u}$~\cite{Hayes2021Multi,Wei2022}. More recent polar Kerr and $\mu$SR measurements on high-quality samples, however, did not observe evidence for TRS breaking below the superconducting critical temperature $T_{c}$~\cite{Ajeesh2023Fate,Azari2023Absence}. To complicate things further, another chiral order parameter, $A_{1u}+iB_{3u}$, has been argued from the temperature dependence of the estimated magnetic penetration depth in tunnel diode measurements~\cite{Ishihara2021Chiral}. Alternatively, $A_{1u}+i\varepsilon B_{3u}$ with a sufficiently small $\varepsilon$ has been suggested by our previous scanning SQUID susceptometry measurements~\cite{Iguchi2023Microscopic}.
 
One possible resolution to these conflicting results is that disorder may play a significant but distinct role in each experiment. 
For example, multiple superconducting phase transitions were reported in UTe$_2$ in specific heat measurements even at ambient pressure~\cite{Hayes2021Multi, Thomas2020}. The splitting of $T_c$ is theoretically expected in an orthorhombic lattice system with multicomponent order parameters $\Gamma_1+i\Gamma_2$ (or $\Gamma_1 +\Gamma_2$ ), where $\Gamma_i$ is an irreducible representation of the point group~\cite{SigristUeda1991}. Subsequently, however, it was found that the double peak in specific heat arises from sample inhomogeneity~\cite{Thomas2021Spat}. Importantly, a single superconducting transition is reported in higher quality samples probed by different techniques, including specific heat ~\cite{Rosa2021Single,Sakai2022Single}, scanning SQUID susceptometry~\cite{Iguchi2023Microscopic}. Accidental degenerate orders were also ruled out by the specific heat measurements under uniaxial stress, which did not cause a splitting of the superconducting transition in single-transition samples~\cite{Girod2022Thermo}. Additionally, recent ultrasound measurements rule out multicomponent order parameters in both single- and double-transition samples~\cite{Theuss2023Single}. 

To address the controversy of whether chiral superconductivity is realized in UTe$_2$, local magnetic imaging of high-quality crystals of UTe$_2$ is key. Here, we imaged the local magnetic flux in the vicinity of edges on the cleaved (011)-plane of single-transition samples of UTe$_2$ at low magnetic fields using scanning SQUID magnetometry. 
We found localized magnetic fields near edges that persisted at low fields and disappeared above $T_c$. The amplitude of the observed edge fields is consistent with the phenomenologically estimated amplitude of the chiral edge current fields.

\section{Methods}
Bulk single crystals of UTe$_2$ were grown by chemical vapor transport. Sample no. 1 and 2 used in this paper (Fig.~S5) were obtained from the same batch as sample S2 in Ref.~\cite{Rosa2021Single}. Heat capacity measurements confirmed a single transition at 1.68 K with a width of 50 mK on a single crystal, which was subsequently cleaved for the two samples used in this study. These two samples were glued on the same copper sample stage. We cooled down the samples twice from room temperature and collected data from sample no.1 on the first and second cool down and no.2 on the second cool down.
We used a scanning SQUID susceptometer to obtain the local ac susceptibility on the cleaved (011)-plane of UTe$_2$ at temperatures varying from 80 mK to 2 K using a Bluefors LD dilution refrigerator. Our scanning SQUID susceptometer had two pairs of a pickup loop and a field coil configured with a gradiometric structure \cite{Kirtleyrsi2016}. The inner diameter of the pickup loop and field coil are 0.8 \textmu m  and 3.0 \textmu m, respectively. The scan height is $\sim$ 0.5 \textmu m. The pickup loop provides the local DC magnetic flux $\Phi$ in units of flux quantum $\Phi_0=h/2e$, where $h$ is the Planck constant. The pickup loop also detects the ac magnetic flux $\Phi^{ac}$ in response to the ac magnetic field $H_0e^{i\omega t}$, which was produced by an AC of $|I^{ac}| =$ 1~mA at frequency $\omega/2\pi\sim$ 1 kHz through the field coil, using an SR830 lock-in amplifier. Here, we report the local dc magnetic flux $\Phi$ and the local ac susceptibility $\chi=\Phi^{ac}/|I^{ac}|$ in units of $\Phi_0$/A.

\begin{figure*}[tb]
\begin{center}
\includegraphics*[width=14cm]{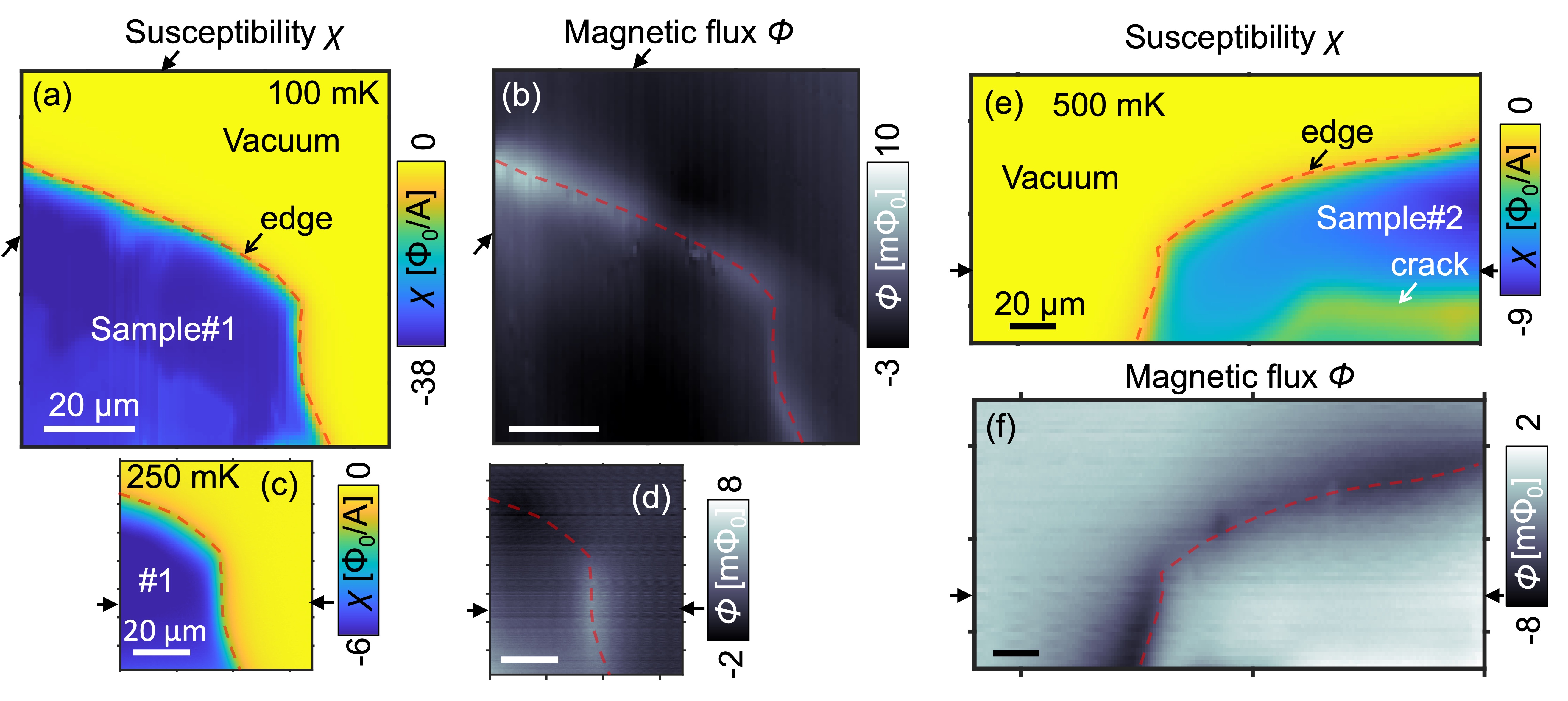}
\caption{{ Imaging local magnetic field distribution along the sample edges of superconducting UTe$_2$ near zero background field.} (a,c,e) Scanning SQUID susceptometry scan visualize crystal edges of (a,c) sample no. 1 and (e) no. 2 below $T_c$. (b,d,f) Scanning SQUID magnetometry scan detects local magnetic field distribution near edges of (b) no. 1 at the first thermal cycle, (d) no. 1 at the second thermal cycle, (f) no. 2 at $H_{\perp}=7.6$~mOe normal to the surface. The difference in signal magnitude at the two samples is due to the different scan heights. Pairs of black arrows indicate the line cuts in Figs. 3,4, S6. Red dashed lines are guides to the eye.}
\end{center}
\end{figure*}

\section{Results}
To investigate the spontaneous magnetism in UTe$_2$, we cooled down two samples through $T_c$  with applied fields that cancel the background fields, which reduces the Meissner screening fields near the edge (Fig. 1c). We used small magnetic fields ($|H|=0-19$~mOe $\ll H_{c1}$), where the lower critical field $H_{c1}\sim$40~mOe at 500 mK~\cite{Ishihara2023Anisotropic}, via a hand-made coil (for details, see the Supplemental Materials~\cite{supple}). Due to its configuration, we note that the hand-made coil produces out-of-plane fields $H_{\perp}$ with in-plane fields $H_{\parallel}$ simultaneously, where  $H_{\parallel}/H_{\perp}\sim$0.4.
We measured the local susceptibility to visualize the sample edges and local magnetic flux in different thermal cycles (2~K $\rightarrow$ 100~mK). We detected a homogeneous single superconducting phase transition from the susceptometry scans in both samples (for details, see Ref.~\cite{Iguchi2023Microscopic}). We successfully visualized the sample shape -- the susceptibility scan of sample no. 1 [no. 2] in Fig.~2a [Fig.~2e] matches the optical microscope image shown in Fig.~S5a [Fig.~S5b]. 

While measuring the local susceptibility, we also measured the local magnetic flux simultaneously.
The observed magnetic flux is reasonably uniform, either inside or outside the sample far from the edge (Fig.~2b). The magnetic flux amplitude inside the sample is the same as outside in Fig. 2b, indicating the background field is almost zero, with a cancellation field  $H_{\perp}$ of $\sim$6~mOe. However, there is apparent local magnetic field distribution near the sample edge. We measured this local edge field three times at 100~mK after heating sample no. 1 to $T = 2~$K $> T_c$, but the sign and profile of this local edge field did not change after thermal cycling. Exposure to air between the first and the second cooling down from room temperature created a dead oxide layer of $\sim$0.7 \textmu m thickness at the surface~\cite{Iguchi2023Microscopic}. As a result, the susceptibility signal weakened during the second thermal cycle. 
Surprisingly, we observed similar edge fields, which suggests this phenomenon is robust against surface oxidization, but the edge fields appear along only one edge (Figs.~2d). In contrast, the susceptibility image resembles the image from the first thermal cycle except for its reduced amplitude (Figs.~2c). We also observed edge fields in sample no. 2 on the second cool down (Figs.~2e,2f), similar to sample no.1 at the first thermal cycle (Fig. 2b). 
We did not see any step-like magnetic anomalies near the sample edges or terraces at zero fields, as seen in Ref.~\cite{Hicks2010Limits} and attributed to the SQUID-sample interaction. We also did not see such anomalies on a reference sample (Fig.~S8) using the same measurement configuration with UTe$_2$ measurements. Therefore, this difference may be due to using a different sensor design.

We also measured local magnetic flux around the sample edges after cooling in different small magnetic fields from 2~K. 
By applying magnetic fields normal to the sample surface, the magnetic flux profile changed, as shown in Fig.~3b (sample no. 1),  Fig.~3e(no. 2), and Fig.~S6b (no. 1). 
After subtracting the zero-field profile, $\Delta\Phi(_{\perp}) = \Phi(H_{\perp}) - \Phi$($\sim$6~mOe), the magnetic flux inside the sample develops to expel the applied fields [sample no. 1 (Fig.~3c) and no.2 (Fig.~3f)], which is consistent with the Meissner screening effect [Fig.~1c] (see also Fig.~S6c (no. 1) in ~\cite{supple}). There are no obvious edge fields in $\Delta\Phi(_{\perp})$, indicating that the amplitude of local magnetic field distribution near the edge does not change by applying the $H_{\perp}$ of 15.5~mOe and $H_{\parallel}$ of 6.9~mOe.  Although it is difficult to calculate the exact magnetic field profile near the superconductor edge in irregular-shaped samples at small background magnetic fields, the edge fields appear robust against small out-of-plane and in-plane background fields.

\begin{figure}[!tb]
\begin{center}
\includegraphics*[width=8cm]{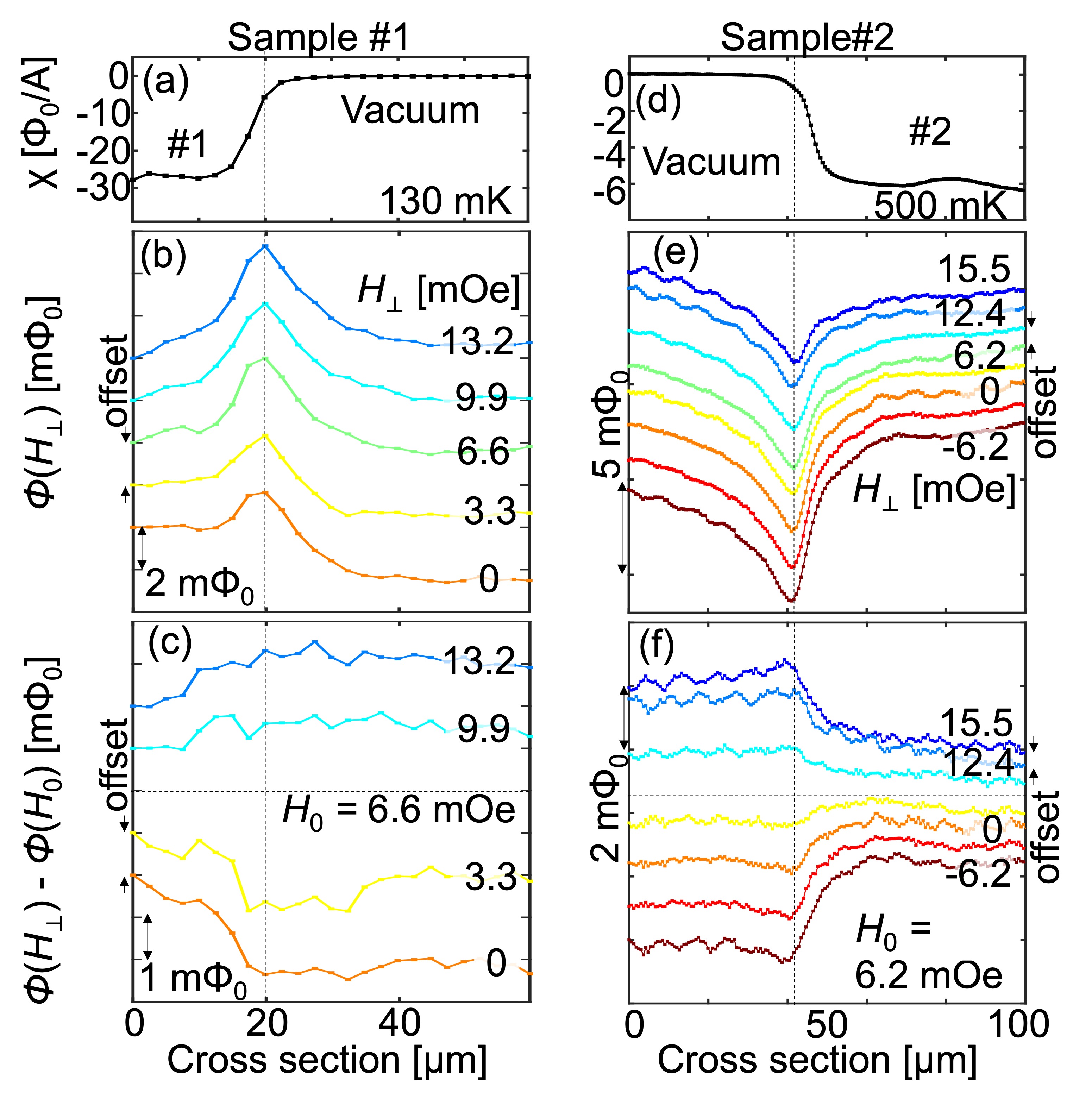}
\caption{{ Magnetic field profile near edges of UTe$_2$ at magnetic background fields.} (a,d) Cross section of susceptometry on (a) sample no. 1 and (d) no. 2. (b,e) Cross section of magnetic flux $\Phi$ near the edge of (b) no. 1 and (e) no. 2 at several applied fields $H_{\perp}$ normal to the scan surface. (c,f) The difference of $\Phi$ at $H_{\perp}$ from $H_0$ (almost zero background out-of-plane fields) shows Meissner screening fields on (c) no. 1 and (f) no. 2. The locations of the cross-section are indicated by pairs of black arrows in Fig. 2.}
\end{center}
\end{figure}

We also observed a similar temperature dependence of the edge fields for our samples within different cool downs. The edge fields disappear above $T_c$ as shown in Figs.~4a and S7a for sample no. 1 and Fig.~4b for no. 2. The peak amplitudes of the edge fields normalized by the data at the base temperature, $T_{\rm min}$ increase rapidly below $T_c$ and saturate at low temperatures (Figs.~4c, S7b). $\Phi_m(T)/\Phi(T_{\rm min})$ looks similar to the behavior of an order parameter. Still, for the quantitative discussion, further theoretical work for the temperature dependence of the chiral edge currents in the possible superconducting state for UTe$_2$ is necessary. 

\begin{figure}[!tb]
\begin{center}
\includegraphics*[width=8cm]{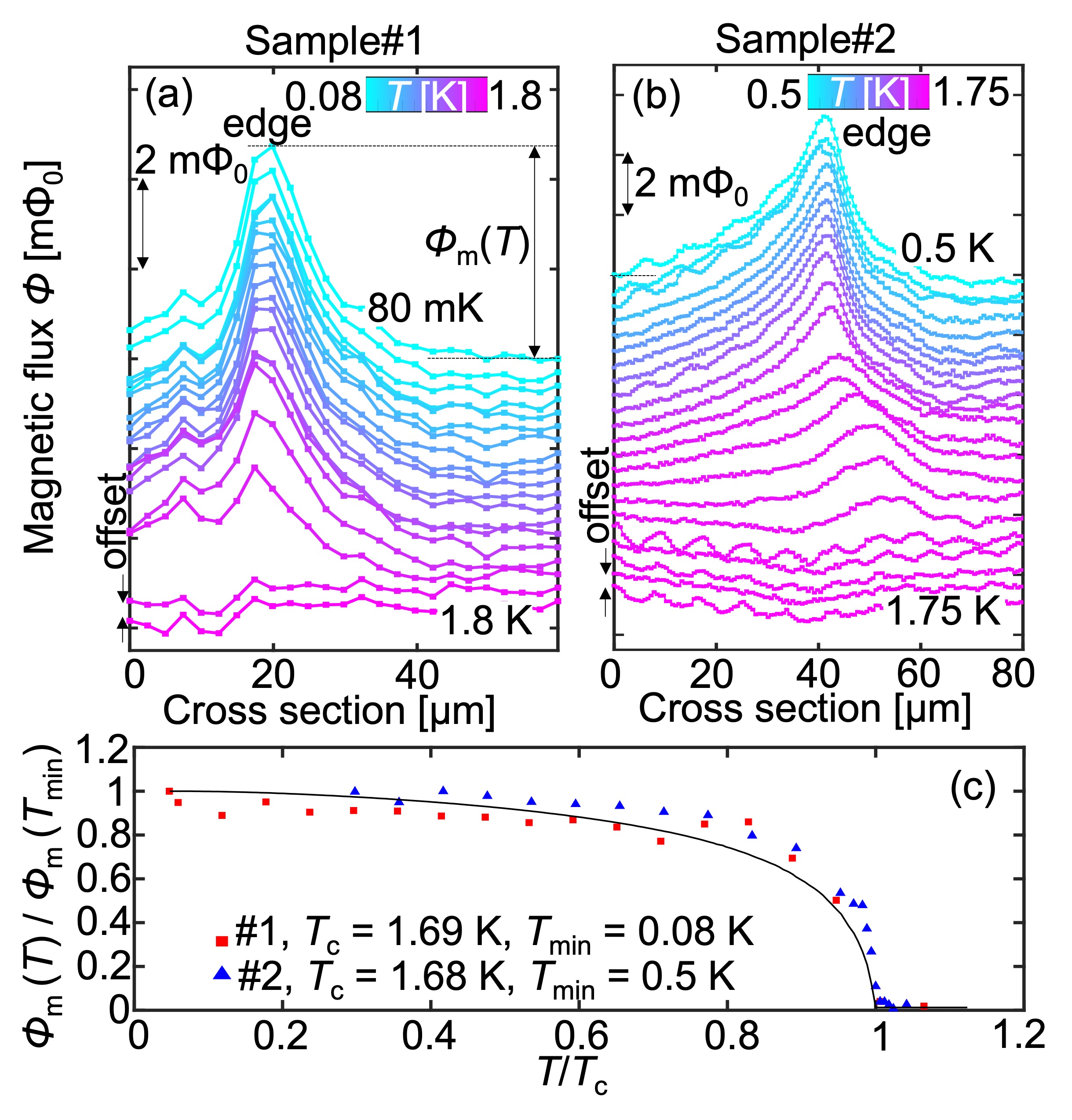}
\caption{{ Edge magnetic field distribution appears in the superconducting phase of UTe$_2$.} (a,b) The temperature dependence of the cross-section of the magnetic flux near the edge of (a) sample no. 1 at $H_0 =$ 6.6~mOe and (b) no. 2 at $H_0 =$ 6.2~mOe. We plot $-\Phi$ in (b) for easy view. (c) Temperature dependence of the normalized peak magnetic flux $\Phi_m(T/T_c)/\Phi_m(T_{min})$ of two samples is almost the same. The difference in signal magnitude at the two samples is due to the different scan heights. The solid line is calculated by using Eq.~(\ref{eq;B-M}).}
\end{center}
\end{figure}

We emphasize that we did not observe any evidence for magnetism in our samples from magnetometry scans above $T_c$ and near zero fields, and a paramagnetic signal was detected over the entire sample in susceptibility scans~\cite{Iguchi2023Microscopic}. We observed sparse weak dipole fields near the edges, as shown in Fig.~S5c (sample no. 1) and Fig.~S5d (no. 2), which may be magnetic impurities. 
The observed edge fields appear robust against these local impurity fields. 
To investigate the background field in our measurement system, scanning SQUID susceptometry and magnetometry measurements were performed on a reference sample in the same system after the first measurements of sample no. 1. There is no obvious local magnetic field distribution near the edges in a reference sample (Fig.~S8).

\section{Discussion}
Our findings reveal the presence of localized magnetic flux in UTe$_2$ below $T_c$ only near the sample edges. Such local magnetic field distribution could be attributed either to the field generated by chiral edge currents or by the Meissner effect resulting from applied magnetic fields or magnetic impurities in the superconducting state. 

First, we discuss the possibility of chiral edge currents. 
We phenomenologically calculate the magnetic flux in a chiral superconductor using the London and Maxwell equations~\cite{supple}.
We obtain the formula of the magnetic flux per unit length along the edge as $\Phi = -(\Phi_0/4)(k_{\rm B}T_c/2\Delta)(\alpha/\xi(\kappa+1))$, where $k_{\rm B}$ is the Boltzmann constant, $\Delta$ is the gap function at zero temperature and $\kappa=\lambda/\xi$. We assume that the unknown parameter $\alpha$ in UTe$_2$ is the same as the model calculation for Sr$_2$RuO$_4$~\cite{Matsumoto1999QuasiparticleSuperconductor}.
We use $\xi = 4.6\sim9.3$ nm~\cite{Niu2020} and $\lambda = 550 \sim 1820$ nm~\cite{Ishihara2021Chiral,Iguchi2023Microscopic} considering that the scanning SQUID picks up the magnetic flux within $\sim 1~$\textmu m length along the edge. We therefore find $|\Phi| = 1.9 \sim 6.4\  {\rm m}\Phi_0$, which is of the same order of magnitude as our experimental results.

In exploring potential order parameters within the Ginzburg-Landau framework for UTe$_2$, we focused on two odd-parity pairing states of the $D_{2h}$ point group. The state $A_u \oplus iB_{1u}$ is pertinent to a (011)-cleaved surface, possibly leading to finite spontaneous magnetization. The interplay of $A_u$ and $B_{1u}$ varies with edge geometry; a detailed analysis of these variations is provided in the supplemental materials. Our findings suggest that the sample’s edge geometry significantly influences spontaneous magnetization, with non-normal edges to the scanning surface potentially inducing magnetization in any chiral symmetries. For precise identification of the chiral axis direction, samples cut into specific geometric shapes are necessary, such as a disk~\cite{Curran2014}.

Next, we consider the extrinsic scenario. We did not detect any chiral domain structures within a $\sim$ 100~\textmu m scale along the edges, which would be expected for chiral superconductivity. The polar Kerr measurements at double-transition sample reported the change of sign or magnitude of the polar Kerr angle, suggesting chiral domain behavior within the beam size of 10~\textmu m~\cite{Hayes2021Multi,Wei2022}. This absence suggests that the observed results could potentially arise from extrinsic factors rather than intrinsic properties of the material.
Here, we discuss two main possible extrinsic scenarios: in-plane magnetic fields (Fig. 1c) and magnetic impurities within the crystals (Fig. 1b). First, the in-plane background field may induce similar edge currents with the chiral edge current. However, our estimations, detailed in the supplemental materials~\cite{supple}, suggest that the in-plane background field required to cause a flux peak of 4~m$\Phi_0$ in a simple rectangular sample would exceed 20~mOe, yet the actual in-plane background field near the sample is approximately 1~mOe at room temperature. Next, no spontaneous magnetization was observed above $T_c$ (Fig.~S5). However, at the phenomenological level, the absence of an observable magnetic source above $T_c$  cannot rule out the possibility of the extrinsic scenario in our noise level~\cite{Bluhm2007Magnetic} (for details, see the Supplemental Materials~\cite{supple}). In a superconductor with ferromagnetic domains, the mean square magnetic flux density $<B_z^2>$ is expressed as
\begin{equation}
\langle B_z^2\rangle = \frac{15\pi}{2}\frac{\lambda^3}{(z+\lambda)^6}\left( \frac{\mu_0}{4\pi}\right)^2M^2 V,
\label{eq;B-M}
\end{equation}
where, $M$ is the magnetization in the domain, $z=0.5~$\textmu m is the scan height, $\mu_0$ is the permeability in vacuum, and $V$ is the volume of the domain. With a temperature-independent $M$, the temperature dependence of the normalized magnetic field $B_z(T)/B_z(0)$ is consistent with the observed edge fields in UTe$_2$ (Fig.~4c), where we use the measured penetration depth~\cite{Iguchi2023Microscopic}.

Our data are consistent with a scenario in which the magnetism arises from a source that is extrinsic to the superconductivity. We consider it to be the most likely scenario in light of recent polar Kerr and $\mu$SR studies, which do not show any evidence of TRSB in the superconducting order parameter~\cite{Ajeesh2023Fate,Azari2023Absence}.
Note that the polar Kerr measurements are, in general, insensitive to the Meissner screening in contrast to scanning SQUID magnetometry~\cite{Wei2022}. 
However if magnetic fields arising from potential magnetic impurities form vortices that are inhomogeneously pinned at local defects, the polar Kerr rotation angle may exhibit changes in the magnetic flux density below $T_c$. Inhomogeneous vortex pinning sites have been observed~\cite{Iguchi2023Microscopic}. At zero temperature, the rotation angle is independent of the applied field, only increasing as the field strength surpasses the point where the vortex-vortex interaction exceeds the pinning potential. This scenario aligns with polar Kerr measurements in double-transition samples~\cite{Hayes2021Multi,Wei2022}. This extrinsic mechanism may also explain our observation of a $\sim$ 100~\textmu m scale edge field, while polar Kerr measurements detected chiral domain behaviour in double-transition samples within 10~\textmu m beam size. 
Previous SQUID measurements indicated a hidden magnetic source in the superconducting state~\cite{Iguchi2023Microscopic}, which is of the same order as our noise level~\cite{supple}, potentially affecting our observations. $\mu$SR and NMR measurements have detected the presence of strong and slow dynamic magnetic fluctuations in UTe$_2$ at low temperatures~ \cite{Tokunaga2022Slow,Sundar2022Ubi}, but, the mechanism by which these dynamic magnetic sources could contribute to static Meissner currents remains unknown and unexplored.

Our results can still be explained by the intrinsic scenario. However, considering these contrasting findings, along with evidence supporting single-component superconductivity in UTe$_2$ at ambient pressure~\cite{Iguchi2023Microscopic,Rosa2021Single,Sakai2022Single,Theuss2023Single}, we propose that the observed edge currents are more likely due to extrinsically induced Meissner currents from hidden magnetic fields. These results underscore the inherent challenges in distinguishing chiral superconductivity through edge magnetic field measurements. Our findings highlight the need for more sensitive methodologies or alternative approaches to overcome these limitations, guiding future research in the quest to accurately characterize chiral superconductivity.

\section{Conclusion}

In conclusion, our scanning SQUID microscopy study reveals localized magnetic fields near the edges of UTe$_2$ near zero field. While the magnitude of these fields aligns with phenomenological calculations for chiral superconductivity, the lack of supporting evidence from recent polar Kerr and $\mu$SR measurements and indications of single-component superconductivity suggest an alternative explanation. 
While these observed edge currents could be extrinsically induced Meissner currents resulting from hidden magnetic fields, this interpretation is not definitive. Our findings underscore the need for further investigation into the complex nature of these hidden fields in UTe$_2$, without conclusively favoring either an intrinsic or extrinsic scenario.

\begin{acknowledgments}
The authors thank C. Hicks, J.R. Kirtley, and Y. Masaki for fruitful discussions. This work was primarily supported by the Department of Energy, Office of Science, Basic Energy Sciences, Materials Sciences and Engineering Division, under Contract No. DE- AC02-76SF00515. Work at Los Alamos was supported by the U.S. Department of Energy, Office of Basic Energy Sciences “Quantum Fluctuations in Narrow-Band Systems” program. H.M. was partially supported by the National Science Foundation under award ECCS-1542152. Y.I. was partially supported by the Japan Society for the Promotion of Science (JSPS), Overseas Research Fellowship. J.I. and M.S. are supported by the Swiss National Science Foundation (SNSF) through Division II (No. 184739).
\end{acknowledgments}



\section*{Contributions}
Y.I. carried out the scanning SQUID microscopy and analyzed data, and wrote the manuscript. H.M. carried out the scanning SQUID microscopy. S.M.T., F.R., and P.F.S.R. synthesized the crystals. J.I and M.S. theoretically calculated the surface magnetization. K.A.M. supervised the project. All the authors discussed the results and implications and commented on the manuscript.
	
	
\section*{Additional information}
Correspondence and requests for materials should be addressed to Y. I. (yiguchi@stanford.edu)



\newpage
\clearpage

\setcounter{figure}{0}
\setcounter{equation}{0}
\renewcommand{\thefigure}{S\arabic{figure}}
\renewcommand{\theequation}{S\arabic{equation}}

\onecolumngrid
\appendix
	
\begin{center}
    \Large
    {Supplemental Material for \\\lq\lq Magnetic edge fields in UTe$_2$ near zero background fields \rq\rq} \\by Iguchi $et$ $al.$
\end{center}

\section{Magnetic flux by chiral edge currents}

To estimate the surface magnetization induced by chiral edge states, we examine a phenomenological theory assuming the spontaneous supercurrent flowing along the ($\bar{1}00$) edge in $y$-direction,
\begin{equation}
    j_y(x) = j_0 e^{-x/\xi},
    \label{eq:driv-curr}
\end{equation}
which decays on a length $\xi$, the coherence length, inside the superconductor ($x >0 $). The induced magnetic field is screened due to the Meissner-Ochsenfeld effect. To take this into account, 
we combine the Maxwell equation $\nabla\times{\bf B}=(4\pi/c){\bf j}$ and the London term $\lambda^{-2}{\bf A}=(4\pi/c){\bf j}$ with the magnetic penetration depth $\lambda$ and obtain
\begin{equation}
    -\partial^2_x A_y(x) + \lambda^{-2} A_y(x) = \frac{4\pi}{c} j_0 e^{-x/\xi},
\end{equation}
where $c$ is the speed of light. Solving this equation by an ansatz:
\begin{equation}
    A_y(x) = a_1 e^{-x/\xi} + a_2 e^{-x/\lambda},
\end{equation}
we find
\begin{equation}
    B_z(x) = \frac{4\pi}{c}j_0\frac{\xi\kappa^2}{\kappa^2-1} \left( e^{-x/\xi}-e^{-x/\lambda} \right).
\end{equation}
Here, $\kappa \equiv \lambda/\xi$ is the Ginzburg-Landau parameter. The magnetic flux per unit length of the edge is
\begin{equation}
    \Phi = \int_{0}^{\infty}{dx B_z(x)} = -\frac{4\pi}{c} j_0 \frac{\xi^2\kappa^2}{\kappa+1}.
\end{equation}
To get a view on the possible magnitude of such a magnetic flux we refer to the discussion of edge currents within a semi-classical approximation for a chiral $p$-wave superconductor\cite{Matsumoto1999QuasiparticleSuperconductor}. 
Under the precondition of specular scattering of electrons at the surface, the self-consistent solution of the Bogoliubov-de Gennes equations leads to an estimate of the current density
\begin{equation}
    j_0 \sim ev_{\rm F} N(\epsilon_{\rm F})k_{\rm B}T_c\alpha,\label{eq:Jy(0)}
\end{equation}
with $\alpha \sim 0.05$ where $e$ is the elementary charge, $v_{\rm F}$ is the Fermi velocity, $N(\epsilon_{\rm F})$ is the electron density of states at Fermi energy, $k_{\rm B}$ is the Boltzmann constant. Although it is important to adapt the situation of UTe$_2$, we expect the order of magnitude to be the same. Therefore, we use Eq.~(\ref{eq:Jy(0)}) for $j_0$ at this stage. By adopting this assumption, we obtain the following formula:
\begin{equation}
    \Phi \sim  -\frac{\Phi_0}{4} \left( \frac{k_{\rm B}T_c}{2\Delta} \right) \frac{\alpha}{\xi(\kappa+1)},\label{eq:Phi_alpha}
\end{equation}
where $\Phi_0= 2\pi\hbar c/2e$ is the magnetic flux quantum, $h = 2\pi\hbar$ is the Planck constant, and $\Delta$ is the gap function at zero temperature. In the derivation of Eq.~(\ref{eq:Phi_alpha}), we used the following standard relations for an isotropic weak-coupling superconductor: $H_c^2/8\pi = N(\epsilon_{\rm F})\Delta^2/2$, $H_{c2} = \sqrt{2}\kappa H_c$, $\xi = \hbar v_{\rm F} / \pi\Delta$, and $H_{c2} = \Phi_0/2\pi\xi^2$. The coherence length $\xi = 4.6\sim9.3$ nm is estimated from the initial slope of $H_{c2}$, where $T_c$ = 1.6~K, $v_{\rm F}$ = 5500$\sim$11000~m/s, $\Delta=1.77k_{\rm B}T_c$ ~\cite{Niu2020}, and the magnetic penetration depth is measured by the previous study $\lambda = 550 \sim 1820$ nm~\cite{Ishihara2021Chiral, Iguchi2023Microscopic}. Considering that the scanning SQUID picks up 1 $\mu$m, we find
\begin{equation}
    |\Phi| = 1.9 \sim 6.4\  {\rm m}\Phi_0,
\end{equation}
which is in the same order as the experimentally observed edge fields.
We note that $|\Phi|$ mainly depends on the penetration depth rather than the coherence length because $\lambda \gg \xi$.

\section{Discussion witin a Ginzburg-Landau theory} 
Using a generalized form of the Ginzburg-Landau theory we
would like to demonstrate here that the edge currents perpendicular to the cleavage direction [011] can be finite if we assume that the order parameter has two odd-parity components and breaks time-reversal symmetry. We consider the representative example of two odd-parity pairing states belonging to the irreducible representations $A_u$ and $B_{1u} $, respectively, of the $D_{2h}$ point group:
\begin{equation}
     {\bf d}_{A_u} ({\bf k}) = c_1k_x\hat{{\bf x}} + c_2k_y\hat{{\bf y}} + c_3k_z\hat{{\bf z}},
\end{equation}
 \begin{equation}
    {\bf d}_{B_{1u}} ({\bf k}) = \bar{c}_1k_y\hat{{\bf x}} + \bar{c}_2k_x\hat{{\bf y}} + \bar{c}_3k_xk_yk_z\hat{{\bf z}},
\end{equation}
where $c_i$, $\bar{c}_i$ are coefficients taking the anisotropy of the orthorhombic system into account. Note that the following discussion can be transferred to a set of other combinations of pairing states. We can assign to each representation a single complex and, generally, position dependent order parameter:
\begin{equation}
    {\bf d}_{A_u} ({\bf k}) \rightarrow \eta_1, \quad {\bf d}_{B_{1u}} ({\bf k}) \rightarrow \eta_2.
\end{equation}
With this simplification, the generalized free energy functional can be written as 
\begin{equation}
    F = \int{d^3r}f(\eta_1,\eta_2),
\end{equation}
with
\begin{eqnarray}
    f &=& a_1|\eta_1|^2 + b_1|\eta_1|^4 + a_2|\eta_2| + b_2|\eta_2|^4 + b_3|\eta_1|^2|\eta_2|^2 + \frac{b_4}{2}(\eta_1^{*2}\eta_2^{2} + \eta_1^{2}\eta_2^{*2} ) \nonumber\\
    && +\sum_{\mu=x,y,z}{\left[ K_{1\mu}|\Pi_\mu\eta_1|^2 + K_{2\mu}|\Pi_\mu\eta_2|^2\right]} + \frac{({\bf \Delta} \times {\bf A})^2}{8\pi}\nonumber\\
    && + K_3\left[ (\Pi_x\eta_1)^*(\Pi_y\eta_2) + (\Pi_x\eta_1)(\Pi_y\eta_2)^*\right]\nonumber\\
    && + K_4 \left[ (\Pi_y\eta_1)^*(\Pi_x\eta_2) + (\Pi_y\eta_1)(\Pi_x\eta_2)^* \right],\label{eq:f}
\end{eqnarray}
where ${\bf \Pi} = (\hbar/i){\bf \nabla} - (2e/c){\bf A}$. This is the most general form of the given order, which is scalar under all symmetries. We assume that the critical temperature of both order parameters is roughly the same and that $b_4>0$ is such that a relative phase of $\pm\pi/2$ is favored between them, leading to a time-reversal symmetry breaking combination. The coefficients $b-i$ have to be chosen to guarantee that the free energy has a stable minimum, and all coefficients $K_i$ are taken positive. 

The current density is obtained by the variation of $f$ with respect to ${\bf A}$,
\begin{equation}
    {\bf j} = -c\frac{\partial f}{\partial {\bf A}}.
\end{equation}
For the spontaneous edge currents, only the mixed gradient terms with coefficient $ K_3 $ and $K_4$ in Eq.~(\ref{eq:f}) are relevant and lead to
\begin{eqnarray}
    j_x &=& -2e\left[ K_3\eta_1^*\Pi_y\eta_2 + K_4\eta_2^*\Pi_y\eta_1 + {\rm c.c.}\right]\nonumber\\
    &=& 2e\hbar i \left[ K_3\eta_1^*\partial_y\eta_2 + K_4\eta_2^*\partial_y\eta_1 - {\rm c.c.}\right],
\end{eqnarray}
\begin{eqnarray}
    j_y &=& -2e\left[ K_3\eta_2^*\Pi_x\eta_1 + K_4\eta_1^*\Pi_x\eta_2 + {\rm c.c.}\right]\nonumber\\
    &=& 2e\hbar i \left[ K_3\eta_2^*\partial_x\eta_1 + K_4\eta_1^*\partial_x\eta_2 - {\rm c.c.}\right],
\end{eqnarray}
where we eventually ignore the vector potential for simplicity since this expression is sufficient to describe the driving currents without Meissner screening. It is obvious that the spontaneous supercurrents only appear if the order parameters have a spatial dependence transverse to the current direction, unlike the usual phase gradient. This behavior is likely to appear near the sample edges due to the pair breaking effects of surface scattering. 

Let us now look at a surface that is perpendicular to the cleavage direction ${\bf n}_c$ = [011], which has normal vector ${\bf n}_s$ with ${\bf n}_s\cdot{\bf n}_c = 0$ whereby the currents run parallel to the surface, ${\bf n}_j$ in a way that
\begin{equation}
    {\bf n}_j\cdot {\bf n}_s = 0,\qquad {\bf n}_j\cdot {\bf n}_c = 0.
\end{equation}
Moreover, both order parameters shall have a special dependence parallel to ${\bf n}_s$ such that
\begin{equation}
    \partial_\mu\eta_{1,2} = n_{s\mu}\partial\eta_{1,2},
\end{equation}
where $\partial$ denotes the derivative $\partial_{x'}$ with $x' = {\bf r}\cdot{\bf n}_s$. From this restriction, the current density parallel to the surface has the form,
\begin{eqnarray}
    j_\parallel &=& {\bf j}\cdot{\bf n}_j\nonumber\\
    &=& 2e\hbar i \left[ n_{jx}n_{sy}(K_3\eta_1^*\partial\eta_2 + K_4\eta_2^*\partial\eta_1) + n_{jy}n_{sx}(K_3\eta_2^*\partial\eta_1 + K_4\eta_1^*\partial\eta_2) - {\rm c.c.} \right]\nonumber\\
    &=& 2e\hbar i \left[ (K_3n_{jx}n_{sy} + K_4n_{jy}n_{sx})\eta_1^*\partial\eta_2 + (K_4n_{jx}n_{sy} + K_3n_{jy}n_{sx})\eta_2^*\partial\eta_1 - {\rm c.c.} \right].
    \label{eq;cur-dens}
\end{eqnarray}

More elaborate discussion of the variational equations shows that to a good approximation, the relative phase between the two components can be fixed to $\pm\pi/2$, leading to 
\begin{equation}
    \eta_1 = g_1(x'),\qquad \eta_2 = sig_2(x'),
\end{equation}
with $s=\pm 1$ denoting the two degenerate time-reversal symmetry breaking phases. Only the order parameter amplitudes vary and should satisfy the condition $
g_j(x') \to |\eta_{j0} | $   
for $ x' \gg \xi $, the mean coherence length, where the order parameter components reach their bulk magnitude $ | \eta_{j0}$ with $ j=1,2$. 

From Eq.(\ref{eq;cur-dens}) we then obtain the simple expression,
\begin{equation}
    j_\parallel = C_1g_1\partial g_2 - C_2g_2\partial g_1,
\end{equation}
with 
\begin{equation}
    C_1 = -4e\hbar s(K_3n_{jx}n_{sy} + K_4n_{jy}n_{sx}),\qquad C_2 = 4e\hbar s (K_4n_{jx}n_{sy} + K_3n_{jy}n_{sx}),
\end{equation}
which only are non-vanishing for broken time-reversal symmetry. 
Because $K_3 \neq K_4$ and $g_1(x') \neq g_2(x')$, in general, we expect a current to flow along the edge for a (011)-cleaved surface for the state $A_u \oplus iB_{1u}$. 

Using a very simple spatial dependence for $ g_j(x')$,
\begin{equation} 
g_j(x') = |\eta_{j0}| (1 - a_j e^{-x'/\xi}) ,
\end{equation}
describing the order parameter reduction close to the surface ($ 0<a_j <1$), 
is sufficient to obtain a form close to the current density of Eq.(\ref{eq:driv-curr}),
\begin{equation}
    j_{\parallel} \approx \frac{|\eta_{10}| | \eta_{20}|}{\xi} ( C_1 a_2 - C_2 a_1 ) e^{-x'/\xi} .
\end{equation}
Note that for geometries with $n_{jx}n_{sy} = n_{jy}n_{sx} = 0$
we do not obtain surface currents. 

\begin{figure*}[htb]
\begin{center}
\includegraphics*[width=7cm]{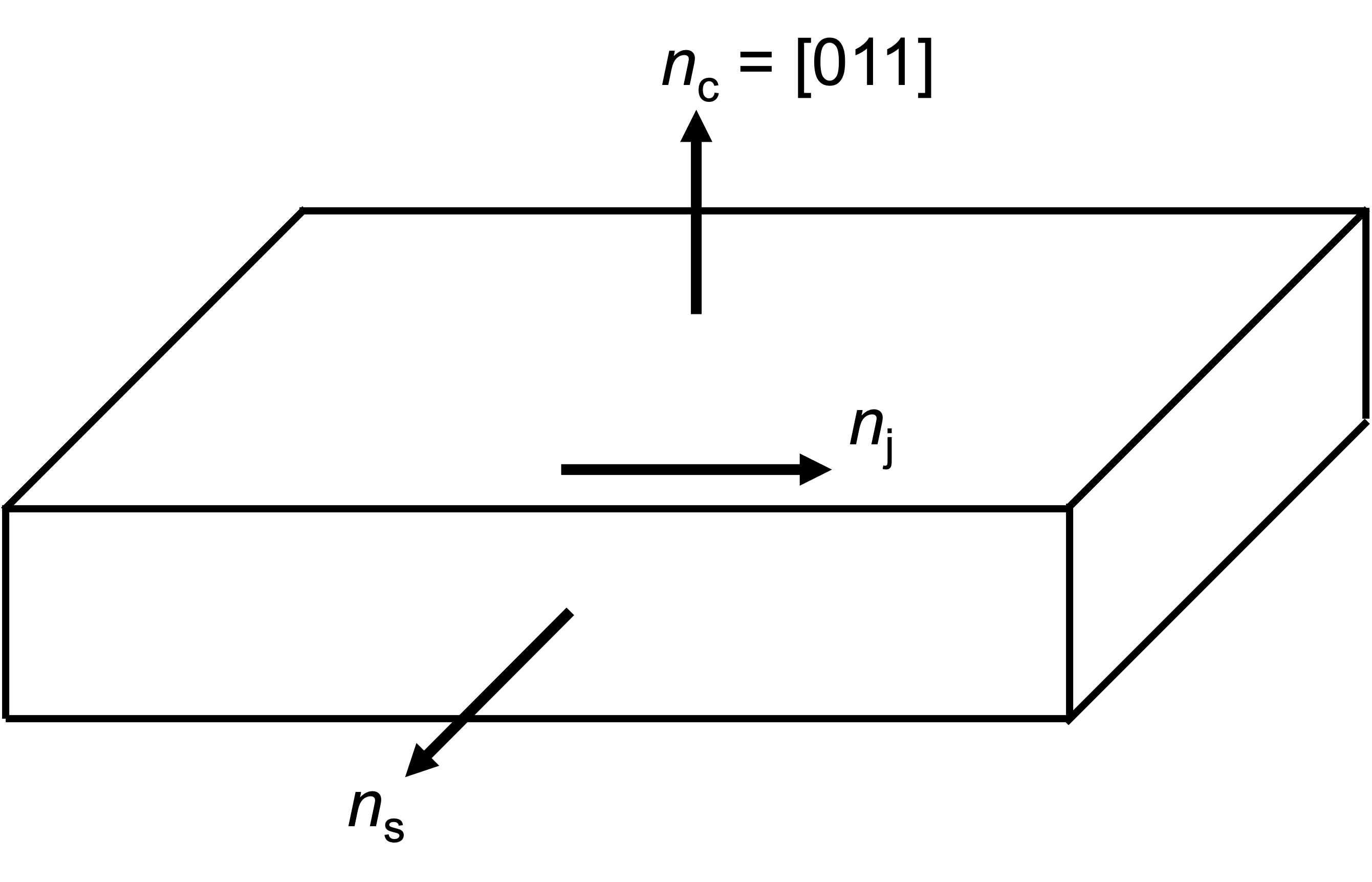}
\caption{Schematic figure of normal vectors.}
\end{center}
\end{figure*}

\clearpage

\section{Magnetic flux by Meissner screening with external fields}
The Meissner screening currents at nonzero external fields also induce the surface magnetic flux. Here, we qualitatively discuss the sign of the surface magnetic fluxes induced by the applied background fields. In addition, we estimate the lower limit of the in-plane background fields that induce the observed edge fields by calculating a simple rectangular-shaped superconductor model and using the experimental results with applied in-plane fields in Fig. 3.

\begin{figure*}[b]
\begin{center}
\includegraphics*[width=12cm]{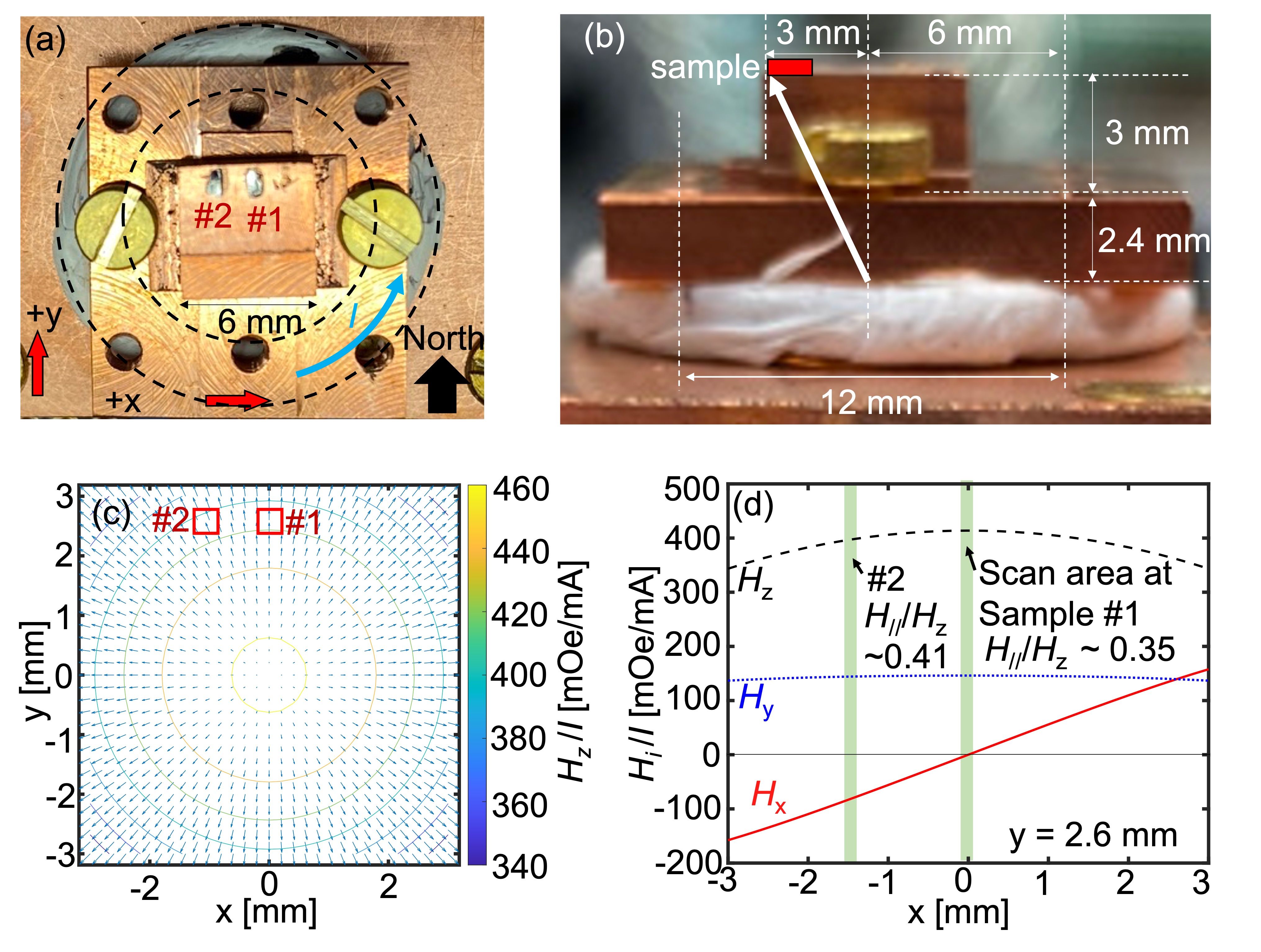}
\caption{Optical microscopy images of (a) the top view and (b) the side view of the copper sample stage with the Nb-Ti hand-made field coil and samples no. 1 and 2. The dashed lines outline the hand-made field coil. (c-e) Simulate 3D Biot-Savart fields divided by the applied current $I$ at the sample stage from the hand-made coil with the dimensions in (b). (c) The allow represents the direction of the in-plane Biot-Savart fields $H_\parallel=(H_x,H_y)$. The counterplot is the out-of-plane Biot-Savart fields $H_z$. (d) The cross-sectional Biot-Savart fields at $y = 2.6$~mm. The green areas imply the scan areas of samples no.1 and 2.}
\end{center}
\end{figure*}

The applied fields $\vec{H}_{ex}$ are not perfectly aligned to the out-of-plane direction. In our measurements, the magnetic fields are applied from the hand-made field coil (Nb-Ti wires, $n =$ 90~turns, inner radius $R =$ 6~mm, see Figs. S2a,b for the configuration details). For this configuration, we numerically simulate the Biot-Savart fields 
\begin{eqnarray}
    H_x (x,y,z) &=& \frac{\mu_0In}{4\pi}Rz\int^{2\pi}_{0}d\theta \cos{\theta} F(x,y,z,\theta),\\
    H_y (x,y,z) &=& \frac{\mu_0In}{4\pi}Rz\int^{2\pi}_{0}d\theta \sin{\theta} F(x,y,z,\theta),\\
    H_z (x,y,z) &=& \frac{\mu_0In}{4\pi}R\int^{2\pi}_{0}d\theta \left( R-x\cos{\theta}-y\sin{\theta} \right) F(x,y,z,\theta), 
\end{eqnarray}
at the sample surface ($z =$ 5.6~mm), where
\begin{equation}
    F(x,y,z,\theta) = \left[ (x-R\cos{\theta})^2+(y-R\sin{\theta})^2+z^2 \right]^{-3/2},
\end{equation}
$\mu_0$ is the
permeability in vacuum, $z=0$ is at the top of the field coil, and 5.6~mm is the total thickness of the sample stage, the sample, grease, and the Teflon tape [Figs. S2c,d]. This simulation provides consistent results with the magnetic fields, estimated by the number of pinned vortices induced by the applied field cooling~\cite{Iguchi2023Microscopic}. The ratio of the in-plane and out-of-plane fields' amplitude $H_\parallel/H_z$ is estimated as 0.35 and 0.41 for sample no.1 and no.2 scan areas, respectively.

\begin{figure*}[b]
\begin{center}
\includegraphics*[width=16cm]{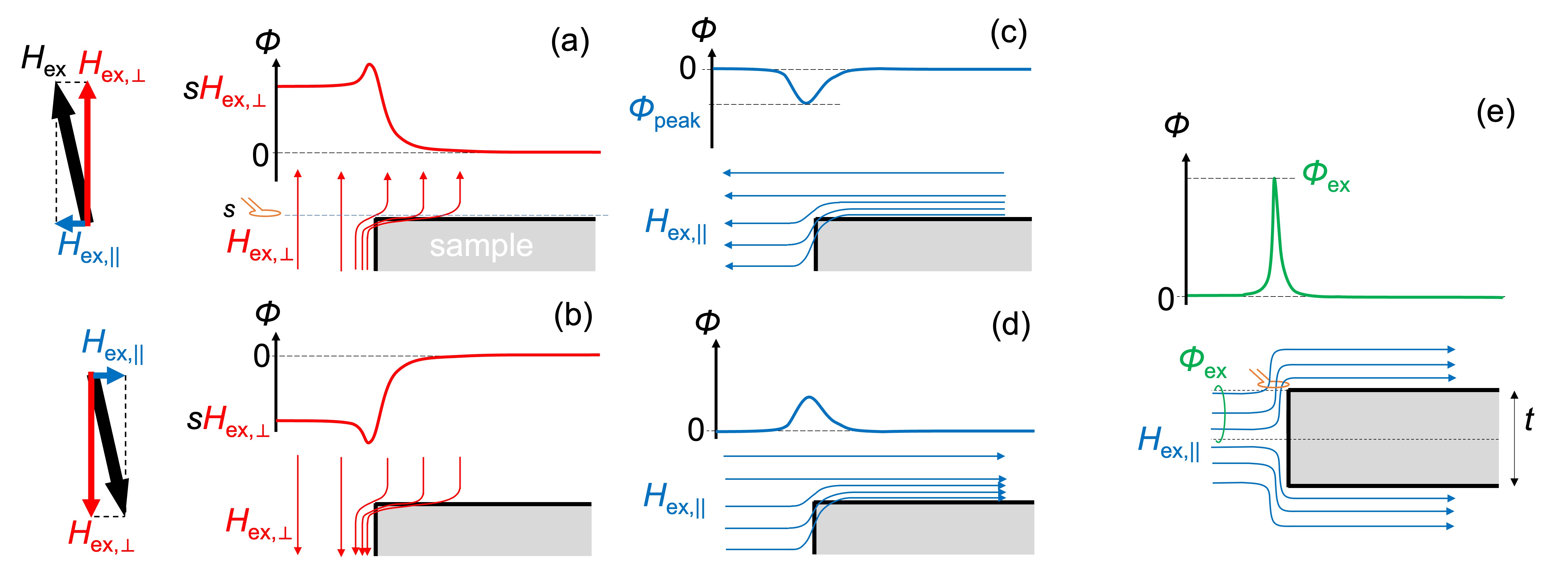}
\caption{Schematic images of magnetic field lines near the sample edge at (a) positive and (b) negative out-of-plane external fields, and at (c) positive and (d) negative in-plane external fields. Insets show the schematic figures of the out-of-plane magnetic flux $\Phi$ along the sample surface. $s$ is the effective area of the pickup loop. (e) A schematic model for overestimating $\Phi_{ex}$ at the edge of a rectangular superconductor with the thickness of $t$ and the in-plane field $H_{ex,\parallel}$.}
\end{center}
\end{figure*}

We consider the Meissner screening for (1) the out-of-plane applied fields $H_{ex, \perp}$ [Figs. S3a \& S3b] and (2) in-plane applied fields $H_{ex,\parallel}$ [Figs. S3c \& S3d]. In case (1), the Meissner currents shield the out-of-plane applied fields near the sample surface. We expect a step-function-like spectrum of the magnetic flux scan along the sample surface, as shown in Figs. S3a and S3b. These spectra are reversed by applying an opposite field, similar to the experimental results in Figs. 2c, 2f, S4d, and S6d. In case (2), the magnetic flux spectrum is expected to peak at the sample edge, as shown in Figs. S3c and S3d. Although the shape of these peaks $\Phi_{peak}$ is similar to what we observed at the edges of UTe$_2$ crystals, these flux spectra should depend on the sign and the amplitude of the applied fields. Therefore, the applied magnetic fields cannot explain our observed edge fields.

Next, we consider the earth's magnetic fields remaining in the sample space, which causes the Meissner currents. We use the cylindrical MuMetal shields of Bluefors, the bottom of which is closed. The background magnetic field around the sample stage at room temperature is observed as $\sim$200~mOe (north, in-plane) and $\sim$300~mOe (out-of-plane) using the commercial fluxgate magnetometer (Mag-03MSES1000, Bartington Instruments). The MuMetal shield reduces the background field at the sample stage to $\sim$1~mOe for both in-plane and out-of-plane components at room temperature. 

To estimate the lower limit of the in-plane external magnetic field ($H_{ex,\parallel}$) that could be the source of the magnetic field observed at the edges of our samples, we build our model based on two assumptions. First, we consider that the in-plane magnetic field applied towards the upper half of the sample's side, which is assumed to have a thin rectangular plate shape, is deflected upward due to the Meissner effect. This effect causes the magnetic field lines to bend perpendicularly away from the surface of a superconductor.

Second, we assume that the resulting deflected magnetic flux ($\Phi_{ex}$) traverses entirely through the pickup loop at the sample's edge. This assumption holds in the context of a sufficiently short magnetic field penetration depth and a negligible distance between the pickup loop and the sample surface, relative to the geometry of the pickup loop. In our model, the pickup loop is simplified as a square. 

In this case, the maximum peak $\Phi_{ex}$ is calculated as $H_{ex, \parallel}lt/2$, where $t\sim100$~\textmu m is the sample thickness and $l\sim$1~\textmu m is the length of one side of a square approximation of the pickup loop. 
Our model suggests that an in-plane magnetic field 1.6~mOe would be required to generate the experimental value of 4~m$\Phi_0$. This magnitude is comparable to the residual external magnetic field observed at room temperature. However, in the actual samples, the magnetic field penetration depth exceeds 1~\textmu m, which is not negligible, and the distance between the pickup loop and the sample surface is about 0.5~\textmu m. Consequently, the magnetic flux peak observed at the edge of the sample is significantly broader -- by an order of magnitude -- than the size of the pickup loop. Therefore, the estimated magnetic flux $\Phi_{ex}$ in our model is considerably larger than it should be. This means that the minimum in-plane magnetic field derived from the model is underestimated. In reality, explaining the experimental results with just the residual external magnetic fields is challenging.

\begin{figure*}[htb]
\begin{center}
\includegraphics*[width=14cm]{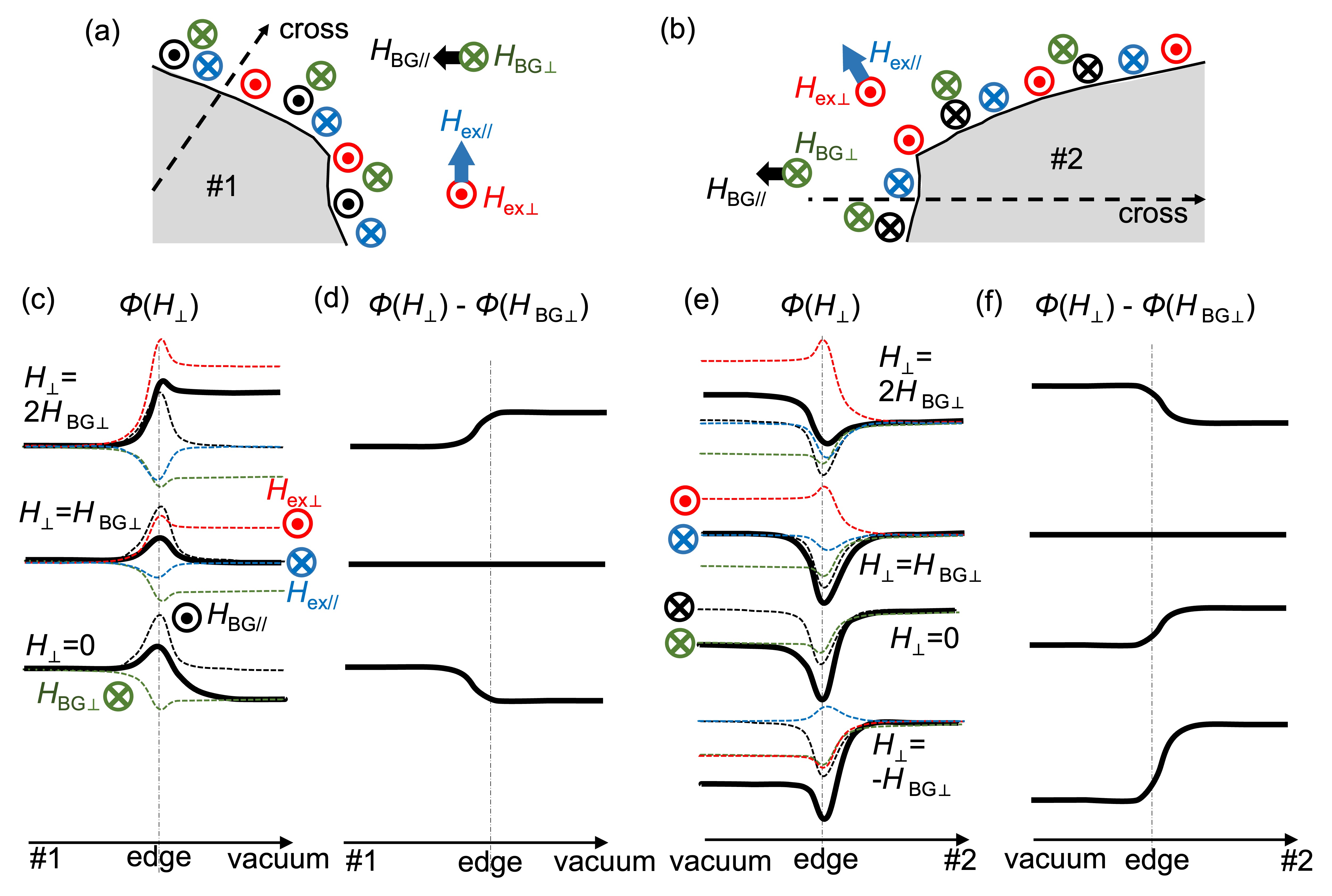}
\caption{{\bf Schematic flux spectra near UTe$_2$ edges with background fields and applied fields.} (a,b) Schematic images of toy models for (a) sample no. 1 and (b) no. 2. The colors of symbols near the edge indicate the sign of the flux peak at the edge due to each field. (c-f) The schematically drawing flux profiles (c),(d) [(e),(f)] for the cross-section in (a) [(b)]. Dashed lines represent fluxes induced by out-of-plane and in-plane components of the background and applied fields.}
\end{center}
\end{figure*}

In our second approach to assess the minimum in-plane magnetic field that generates the edge field, we analyze the experimental data in Fig. 3. Figures S4a and S4b illustrate the initial state of our sample setup, which includes both out-of-plane and in-plane magnetic fields even when no external field is applied.

Specifically, we consider the flux peaks at the sample's edge. If the peak caused by the in-plane component of the background field ($H_{BG\parallel}$) is higher than that caused by the out-of-plane component ($H_{BG\perp}$), the resulting magnetic flux profile would resemble that shown in Figs. 3b and 3e without any applied out-of-plane field, as depicted in Figs. S4c and S4e $H_{\perp}=0$.

When we introduce applied magnetic fields, we observe alterations in the magnetic flux profile, as shown in Figs. 3c and 3f. These changes suggest that the flux peak from the out-of-plane component ($H_{\perp}$) is negated when the in-plane component is approximately 0.4 times the magnitude of the out-of-plane component ($H_{\parallel}\sim0.4H_{BG\perp}$), a relationship supported by our configuration in Figs. S4c to S4f. Note that in our configuration, $H_{\perp}$ induces a flux peak, the sign of which is opposite to that caused by $H_{\parallel}$.

From these observations, we deduce that $H_{BG\parallel}$ must be more substantial than $0.4H_{BG\perp}$. We determine the out-of-plane background fields to be essentially zero when the magnetic flux within the sample matches that outside. This balance is achieved by applying a current of 0.2~mA through our custom field coil, resulting in background fields of 6.6 and 6.2~mOe for samples no. 1 and no. 2, respectively, and corresponding in-plane fields of 2.3 and 2.5~mOe. Thus $H_{BG\perp}$ is estimated as $\sim$6~mOe. This is larger than our room temperature measurements, which might be related to hidden magnetic fields observed in our previous report~\cite{Iguchi2023Microscopic}. 

At $H_\perp = H_{BG\perp}$ in our model, the edge peak is induced by a combination of $H_{BG\parallel}$ and $H_{\parallel}$. At sample no. 1, the peak height caused by $H_{BG\parallel}$ is reduced by $H_{\parallel}=2.3$~mOe. At sample no.2, the peak height caused by $H_{BG\parallel}$ is enhanced by $H_{\parallel}=2.5$~mOe. However, both peaks are $\sim$4~m$\Phi_0$. Therefore, 5~mOe in-plane fields cannot induce significant flux peaks at the edges, indicating the upper limit of the in-plane background field causing 1~m$\Phi_0$ is roughly estimated to be 5~mOe. To produce a peak height of 4~m$\Phi_0$, the lower limit of the in-plane background field is estimated to be 20~mOe. To conclude, the lower bound of the residual in-plane magnetic field estimated here is more stringent than our initial model's estimation and significantly exceeds the measured room-temperature value of $\sim$1~mOe. This finding suggests that the residual in-plane magnetic field model alone cannot account for our experimental results. 

\section{Magnetic flux by magnetic inclusions}
To estimate the possible magnetization due to any magnetic inclusions in our samples, which induces the observed edge fields, we use the phenomenological calculation~\cite{Bluhm2007Magnetic}. When we consider the ferromagnetic domains in a superconductor, the mean square magnetic flux density $<B_z^2>$ in units of T$^2$ is expressed as
\begin{equation}
\langle B_z^2\rangle = \frac{15\pi}{2}\frac{\lambda^3}{(z+\lambda)^6}\left( \frac{\mu_0}{4\pi}\right)^2M^2 V
\end{equation}
where, $M$ is the magnetization in the domain in units of A/m, $z=0.5~\mu$m is the scan height in units of m, $\lambda=550-1820$~nm, $\mu_0=4\pi\times10^{-7}$~N/A$^2$, and $V=L^3$ is the volume of the domain in units of m$^3$. This formula is usually used to estimate the upper limit of the domain size by using the noise of the magnetometry where no edge fields is observed. However, in this study, we observed the finite edge fields. Thus, we calculate the possible domain size and compare it with the sample size. From the experimental noise floor in the magnetometry above $T_c$ is $\sim0.2$~m$\Phi_0$, which is in the same order with the hidden magnetic fields of 0.4-0.8~m$\Phi_0$~\cite{Iguchi2023Microscopic}, we estimate the upper limit of the magnetization of UTe$_2$ as $M=0.41$~A/m, where we use the effective area of the pickup loop of $\pi(500~{\rm nm})^2$. The observed edge fields in UTe$_2$ of 4~m$\Phi_0$ is obtained with $M=0.41$~A/m and $L\geq28-41$~$\mu$m. The obtained domain size is on the same order as the scan area, which is much smaller than the sample size. Therefore, our results of the magnetometry without spontaneous magnetization above $T_c$ in Fig.~S5 cannot rule out the magnetic inclusion scenario.

\newpage

\begin{figure*}[htb]
\begin{center}
\includegraphics*[width=11cm]{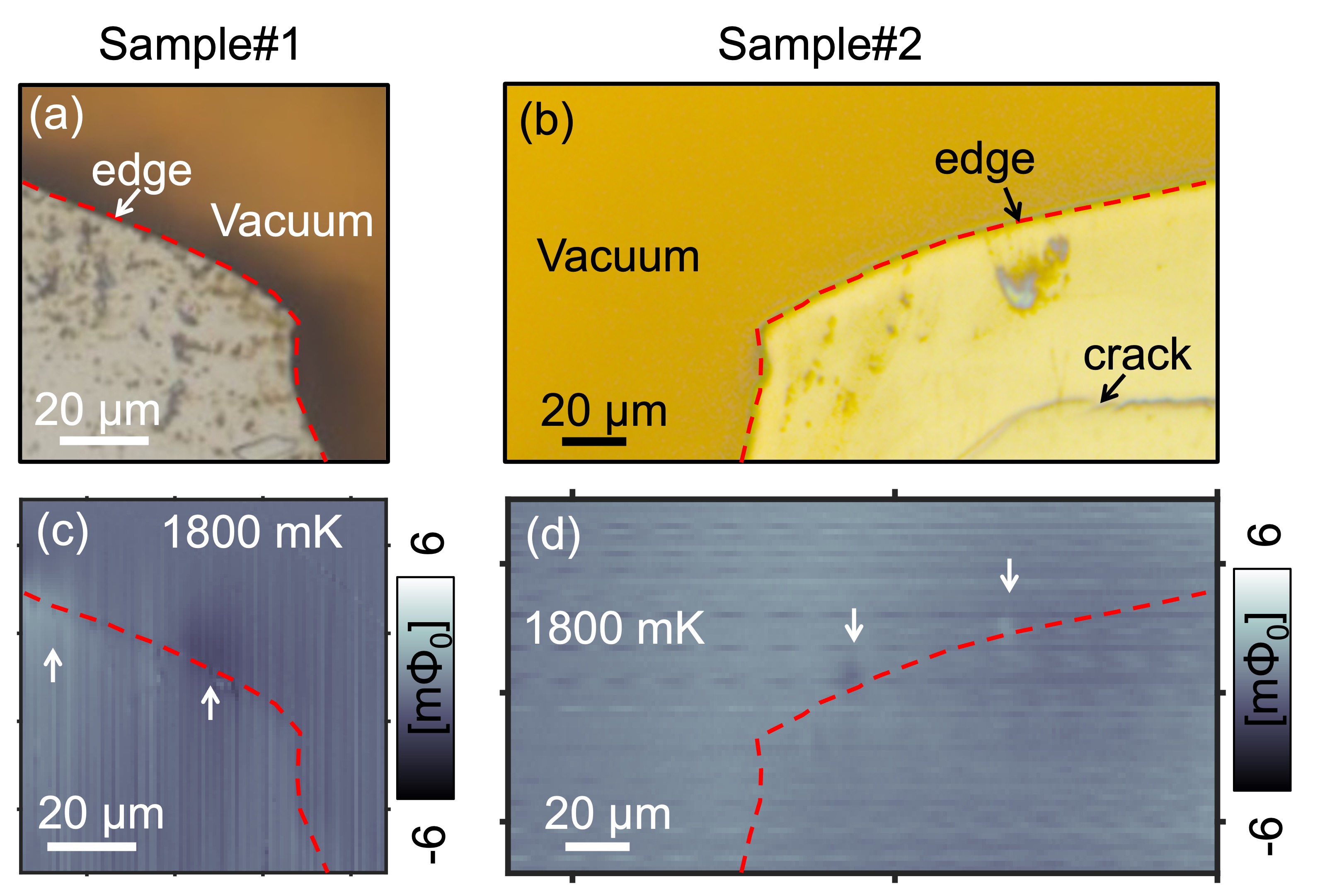}
\caption{{\bf Imaging local magnetic field distribution along the sample edges of paramagnetic UTe$_2$ near zero background field.} (a,b) Optical microscopy images of (a) sample no. 1 and (b) no. 2. (c,d) Above $T_c$ there is no strong magnetic source to induce Meissner screening field near the edges of (c) sample no. 1 and (d) no. 2. Some weak dipole fields indicated by arrows are also visible below $T_c$ in Figs. 2b \& 2f. Red dashed lines are guides to the eye.}
\end{center}
\end{figure*}

\begin{figure*}[htb]
\begin{center}
\includegraphics*[width=5cm]{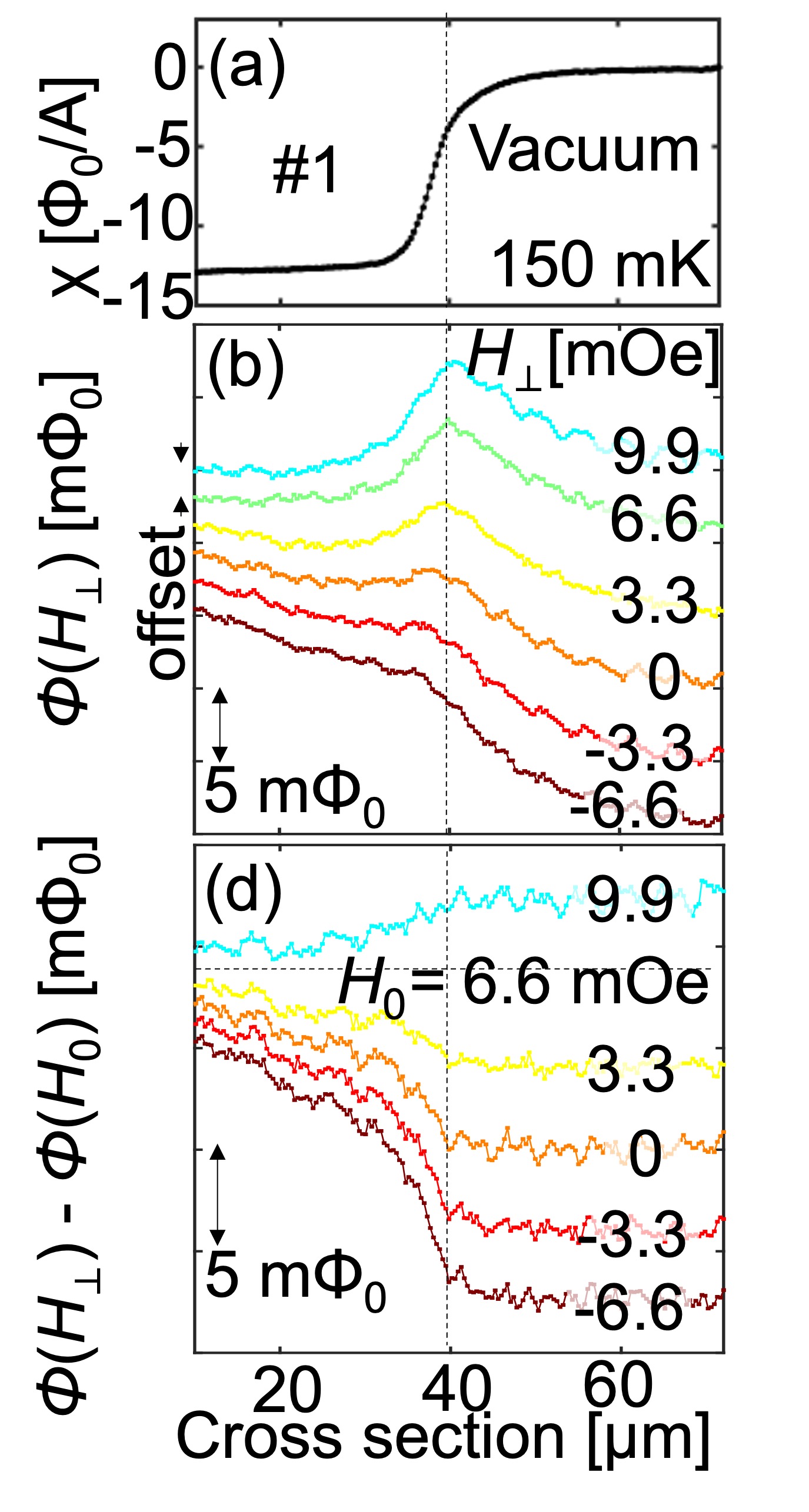}
\caption{ Magnetic field profile near edges of UTe$_2$ sample no. 1 at second thermal cycle at magnetic background fields. (a,c) Cross section of susceptometry on sample no. 1 at second thermal cycle. Plot the same data in (c) with (a). (b) Cross section of magnetic flux $\Phi$ near the edge of sample no. 1 at several background fields $H$ normal to the scan surface. (d) The difference of $\Phi$ at $H$ from $H=0$ shows Meissner screening fields on sample no. 1. The location of the cross-section is indicated by the pair of black arrows in Figs. 2c,2d.}
\end{center}
\end{figure*}

\begin{figure*}[htb]
\begin{center}
\includegraphics*[width=6cm]{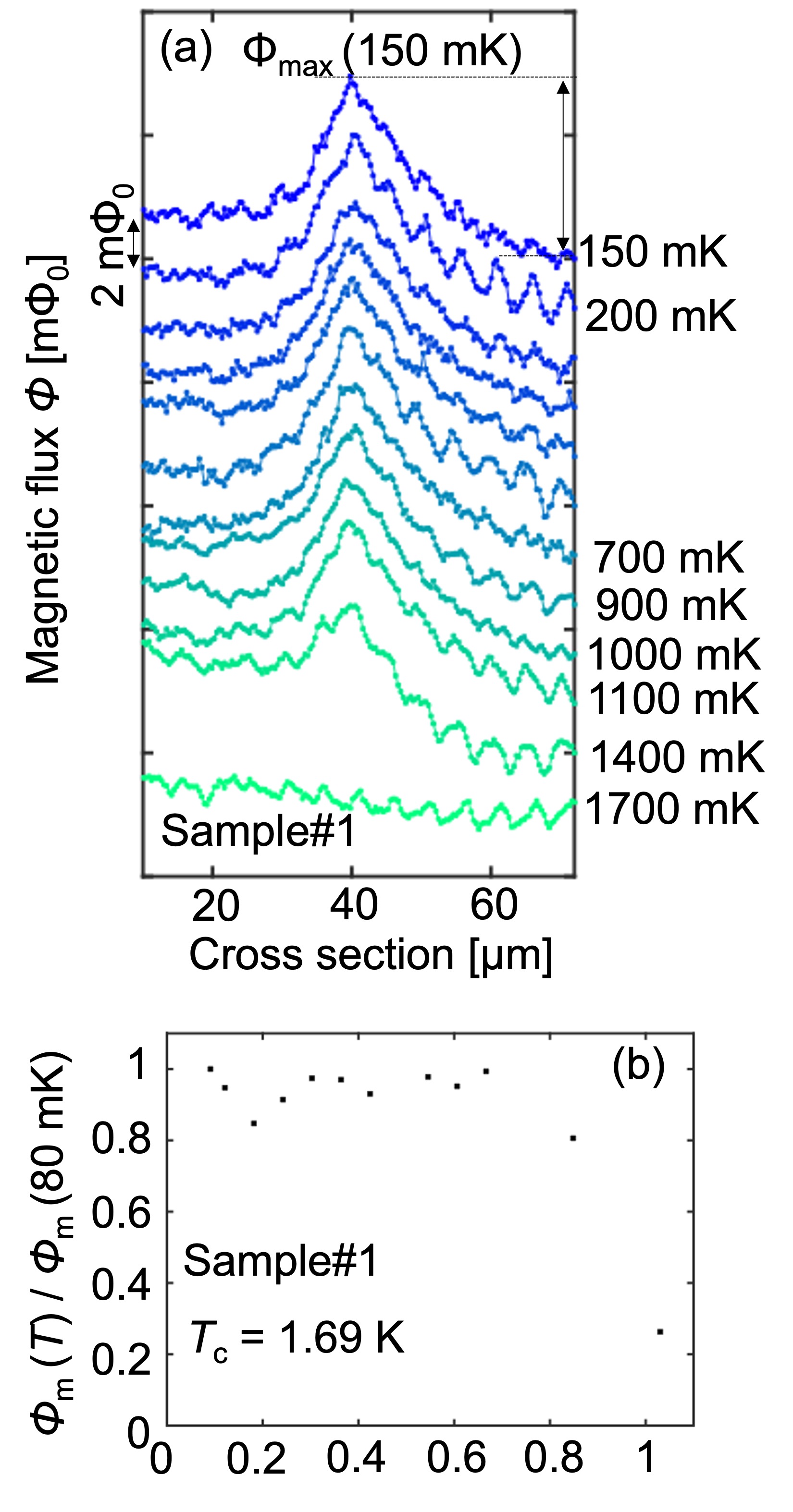}
\caption{Edge magnetic excitations appear in the superconducting phase of UTe$_2$ sample no. 1 at the second thermal cycle. (a) The temperature dependence of the cross-section of the magnetic flux near the edge of sample no. 1 at the second thermal cycle. (b) Temperature dependence of the normalized peak magnetic flux is almost the same as other results in Fig.~4.}
\end{center}
\end{figure*}

\begin{figure*}[htb]
\begin{center}
\includegraphics*[width=10cm]{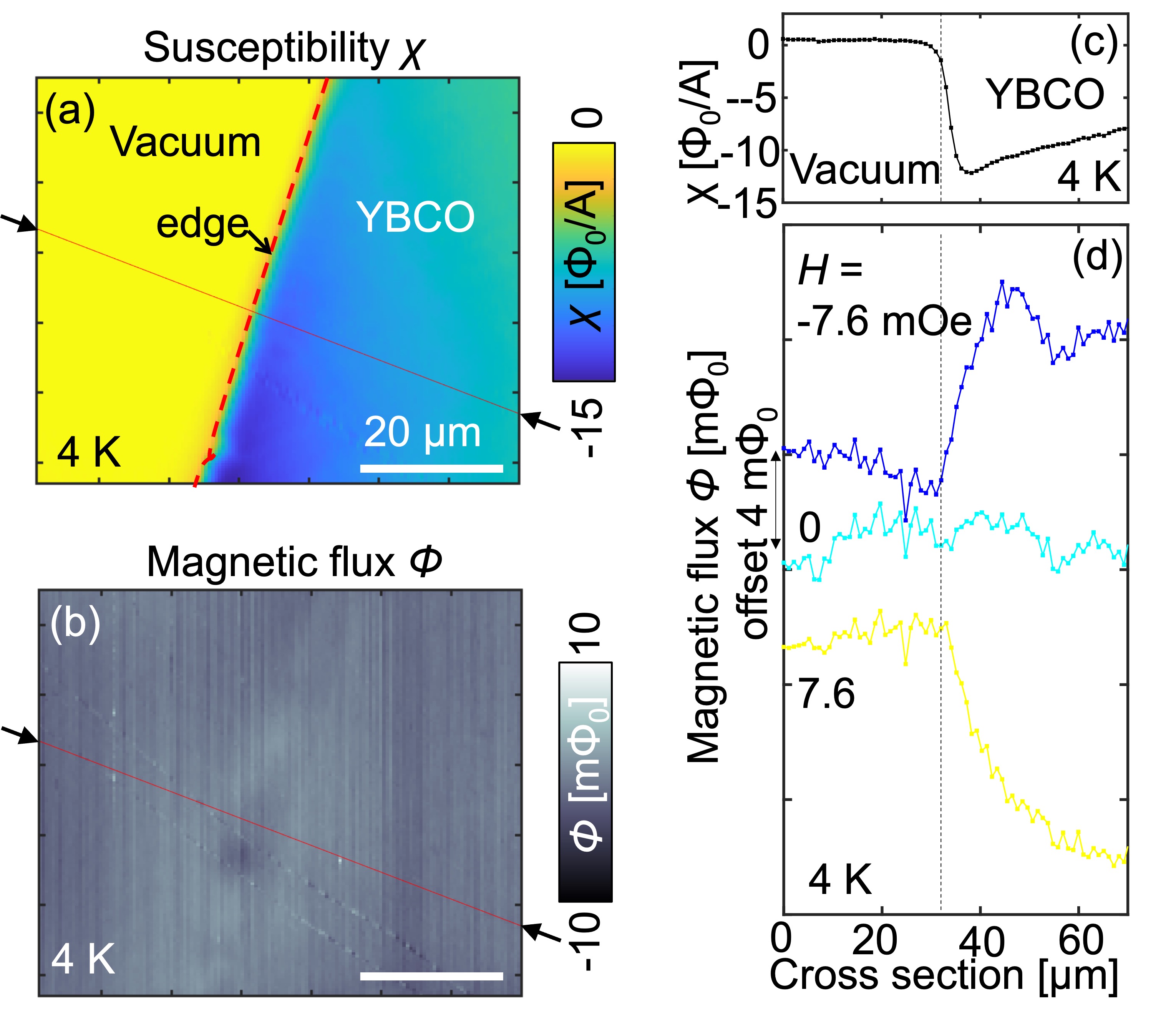}
\caption{No apparent edge field was observed in a reference sample at zero field. (a,b) On the $c$ plane of YBCO (single crystal of high-$T_c$ superconductor YBa$_2$Cu$_3$O$_{6.358}$ with $T_c\sim$17 K)at $H=0$ (a) the susceptibility scan shows the sample edge and (b) the magnetic flux scan has no strong local excitation near edge at 4 K ($T/T_c\sim0.23$). (c,d) The cross sections of (c) susceptibility and (d) magnetic flux at some magnetic fields. The locations of the cross-section are indicated by pairs of black arrows in (a) or (b).}
\end{center}
\end{figure*}


\begin{thebibliography}{}
    \bibitem{Tanaka2012} Y. Tanaka, M. Sato, and N. Nagaosa, J. Phys. Soc. Jpn. {\bf 81}, 011013 (2012).
    \bibitem{Alicea2012NewSystems} J. Alicea, Rep. Prog. Phys. {\bf 75}, 076501 (2012).
    \bibitem{Mizushima2015Symmetry3He-B} T. Mizushima, Y. Tsutsumi, M. Sato, and K. Machida, J. Phys.: Condens. Matter {\bf 27}, 113203 (2015). 
    \bibitem{Nie2020EdgeRevisited} W. Nie, W. Huang, and H. Yao, Phys. Rev. B {\bf 102}, 054502 (2020).
    \bibitem{Matsumoto1999QuasiparticleSuperconductor} M. Matsumoto and M. Sigrist, J. Phys. Soc. Jpn. {\bf 68}, 994 (1999).
    \bibitem{Kasai2018Chiral} J. Kasai, Y. Okamoto, K. Nishioka, T. Takagi, and Y. Sasaki, Phys. Rev. Lett. {\bf 120}, 205301 (2018).
    
    \bibitem{Kallin2016} C. Kallin and J. Berlinsky, Rep. Prog. Phys. {\bf 79}, 054502 (2016).
    \bibitem{Wysokinski2019} K. I. Wysoki\'{n}ski, Condens. Matter {\bf 4}, 47 (2019).
    
    \bibitem{Avers2020Broken} K.E. Avers, W.J. Gannon, S.J. Kuhn {\it et al}., Nat. Phys. {\bf 16}, 531–535 (2020). 
    
    \bibitem{Luke1993} G. M. Luke, A. Keren, L. P. Le, W. D. Wu, Y. J. Uemura, D. A. Bonn, L. Taillefer, and J. D. Garrett, Phys. Rev. Lett. {\bf 71}, 1466 (1993). 
    \bibitem{Schemm2014} E.R. Schemm, W. J. Gannon, C. M. Wishne, W. P. Halperin, and A. Kapitulnik, Science {\bf 345}, 190 (2014).
    
    \bibitem{Luke1998} G. M. Luke {\it et al}., Nature {\bf 394}, 558 (1998).
    \bibitem{Xia2006} J. Xia, Y. Maeno, P. T. Beyersdorf, M. M. Fejer, and A. Kapitulnik, Phys. Rev. Lett. {\bf 97}, 167002 (2006).
    
    \bibitem{Aoki2003} Y. Aoki {\it et al}., Phys. Rev. Lett. {\bf 91}, 067003 (2003).
    \bibitem{Levenson2018} E. M. Levenson-Falk, E. R. Schemm, Y. Aoki, M. B. Maple, and A. Kapitulnik, Phys. Rev. Lett. {\bf 120}, 187004 (2018).
    
    \bibitem{Schemm2015} E. R. Schemm, R. E. Baumbach, P. H. Tobash, F. Ronning, E. D. Bauer, and A. Kapitulnik, Phys. Rev. B {\bf 91}, 140506(R) (2015).
    
    
    \bibitem{Bjornsson2005} P. G. Bj\"{o}rnsson, Y. Maeno, M. E. Huber, and K. A. Moler, Phys. Rev. B {\bf 72}, 012504 (2005).
    \bibitem{Kirtley2007Upper} J. R. Kirtley, C. Kallin, C. W. Hicks, E.-A. Kim, Y. Liu, K. A. Moler, Y. Maeno, and K. D. Nelson, Phys. Rev. B {\bf 76}, 014526 (2007).
    \bibitem{Hicks2010Limits} C. W. Hicks, J. R. Kirtley, T. M. Lippman, N. C. Koshnick, M. E. Huber, Y. Maeno, W. M. Yuhasz, M. B. Maple, and K. A. Moler, Phys. Rev. B {\bf 81}, 214501 (2010).
    \bibitem{Curran2014} P. J. Curran, S. J. Bending, W. M. Desoky, A. S. Gibbs, S. L. Lee, and A. P. Mackenzie, Phys. Rev. B {\bf 89}, 144504 (2014).
    \bibitem{Iguchi2021} Y. Iguchi, I. P. Zhang, E. D. Bauer, F. Ronning, J. R. Kirtley, and K. A. Moler, Phys. Rev. B {\bf 103}, L220503 (2021).
    \bibitem{Curran2023} P.J. Curran, S.J. Bending, A.S. Gibbs, and A. P. Mackenzie, Sci Rep 13, 12652 (2023).

    \bibitem{Bluhm2007Magnetic} H. Bluhm, Phys. Rev. B {\bf 76}, 144507, (2007).
    
    \bibitem{Ran2019Nearly} S. Ran {\it et al}., Science {\bf 365}, 684 (2019).
    \bibitem{Aoki2022Unconventional} D. Aoki, J.-P. Brison, J. Flouquet, K. Ishida, G. Knebel, Y. Tokunaga, and Y. Yanase, J. Phys.: Condens. Matter {\bf 34}, 243002 (2022).
    \bibitem{Nakamine2019Sup} G. Nakamine {\it et al}., J. Phys. Soc. Jpn. {\bf 88}, 113703 (2019).
    \bibitem{Nakamine2021Ani} G. Nakamine {\it et al}., Phys. Rev. B {\bf 103}, L100503 (2021).
    \bibitem{Matsumura2023Large} H. Matsumura {\it et al}., J. Phys. Soc. Jpn. {\bf 92}, 063701 (2023).
    \bibitem{Metz2019Point} T. Metz, S. Bae, S. Ran, I.-L. Liu, Y. S. Eo, W. T. Fuhrman, D. F. Agterberg, S. M. Anlage, N. P. Butch, and J. Paglione, Phys. Rev. B {\bf 100}, 220504(R) (2019).
    
    \bibitem{Kittaka2020Ori} S. Kittaka, Y. Shimizu, T. Sakakibara, A. Nakamura, D. Li, Y. Homma, F. Honda, D. Aoki, and K. Machida, Phys. Rev. Research {\bf 2}, 032014(R) (2020).
    \bibitem{Lee2023Ani} S. Lee, A. J. Woods, P. F. S. Rosa, S. M. Thomas, E. D. Bauer, S.-Z. Lin, and R. Movshovich, arXiv:2310.04938.
    \bibitem{Bae2021Ano} S. Bae, H. Kim, Y. S. Eo, S. Ran, I-l. Liu, W. T. Fuhrman, J. Paglione, N. P. Butch, and S. M. Anlage, Nat. Commun. {\bf 12}, 2644 (2021).
    \bibitem{Ishihara2021Chiral} K. Ishihara, M. Roppongi, M. Kobayashi, Y. Mizukami, H. Sakai, Y. Haga, K. Hashimoto, and T. Shibauchi, Nat. Commun. {\bf 14}, 2966 (2023).
    \bibitem{Iguchi2023Microscopic} Y. Iguchi, H. Man, S. M. Thomas, F. Ronning, P. F. S. Rosa, and K. A. Moler, Phys. Rev. Lett. {\bf 130}, 196003 (2023).
    \bibitem{Jiao2020Chiral} L. Jiao, S. Howard, S. Ran, Z. Wang, J. O. Rodriguez, M. Sigrist, Z. Wang, N. P. Butch, and V. Madhavan, Nature {\bf 579}, 523 (2020).
    \bibitem{Hayes2021Multi} I. M. Hayes {\it et al}., Science {\bf 373}, 797 (2021).
    \bibitem{Wei2022} D. S. Wei {\it et al}., Phys. Rev. B {\bf 105}, 024521 (2022).
    \bibitem{Ajeesh2023Fate} M.O. Ajeesh, M. Bordelon, C. Girod, S. Mishra, F. Ronning, E.D. Bauer, B. Maiorov, J.D. Thompson, P.F.S. Rosa, and S.M. Thomas, Phys. Rev. X {\bf 13}, 041019 (2023).
    \bibitem{Azari2023Absence} N. Azari, M. Yakovlev, N. Rye, S. R. Dunsiger, S. Sundar, M. M. Bordelon, S. M. Thomas, J. D. Thompson, P. F. S. Rosa, and J. E. Sonier, Phys. Rev. Lett. {\bf 131}, 226504 (2023). 
    \bibitem{Thomas2020} S. M. Thomas, F. B. Santos, M. H. Christensen, T. Asaba, F. Ronning, J. D. Thompson, E. D. Bauer, R. M. Fernandes, G. Fabbris, and P. F. S. Rosa, Sci. Adv. {\bf 6}, eabc8709 (2020).
    \bibitem{SigristUeda1991} M. Sigrist and K. Ueda, Rev. Mod. Phys. {\bf 63}, 239 (1991).
    \bibitem{Thomas2021Spat} S. M. Thomas, C. Stevens, F. B. Santos, S. S. Fender, E. D. Bauer, F. Ronning, J. D. Thompson, A. Huxley, and P. F. S. Rosa, Phys. Rev. B {\bf 104}, 224501 (2021).  

    \bibitem{Rosa2021Single} P. F. S. Rosa, A. Weiland, S. S. Fender, B. L. Scott, F. Ronning, J. D. Thompson, E. D. Bauer, and S. M. Thomas, Commun Mater {\bf 3}, 33 (2022).
    \bibitem{Sakai2022Single} H. Sakai, P. Opletal, Y. Tokiwa, E. Yamamoto, Y. Tokunaga, S. Kambe, and Y. Haga, Phys. Rev. Materials {\bf 6}, 073401 (2022)
    \bibitem{Girod2022Thermo} C. Girod {\it et al}., Phys. Rev. B {\bf l06}, L121101 (2022).
    \bibitem{Theuss2023Single} F. Theuss {\it et al}., Nat. Phys. {\bf 20}, 1124 (2024). 
    
    \bibitem{Kirtleyrsi2016} J. R. Kirtley {\it et al}., Rev. Sci. Instrum. {\bf 87}, 093702 (2016).

    \bibitem{Ishihara2023Anisotropic} K. Ishihara , M. Kobayashi, K. Imamura, M. Konczykowski, H. Sakai, P. Opletal, Y. Tokiwa, Y. Haga, K. Hashimoto, and T. Shibauchi, Phys. Rev. Res. {\bf 5}, L022002 (2023).
    
    
    \bibitem{supple} See Supplemental Material at [URL will be inserted by publisher] for the details of the phenomenology of surface magnetization.
    
    \bibitem{Niu2020} Q. Niu, G. Knebel, D. Braithwaite, D. Aoki, G. Lapertot, G. Seyfarth, J.-P. Brison, J. Flouquet, and A. Pourret, Phys. Rev. Lett. {\bf 124}, 086601 (2020).
    
    \bibitem{Tokunaga2022Slow} Y. Tokunaga {\it et al.}, J. Phys. Soc. Jpn. {\bf 91}, 023707 (2022).
    \bibitem{Sundar2022Ubi} S. Sundar {\it et al.}, Commun. Phys. {\bf 6}, 24 (2023). 	
    
\end{thebibliography}
\end{document}